\documentclass[a4paper,12pt]{article}

\usepackage[left=2.5cm,right=2.5cm,top=2.5cm,bottom=2.5cm]{geometry}
\usepackage{graphicx}
\usepackage{amssymb}
\usepackage{multirow}
\usepackage{amsmath}
\usepackage{psfrag}
\usepackage{setspace}
\usepackage{subcaption}
\usepackage[colorlinks=true]{hyperref}
\usepackage{float}
\usepackage{cite}

\hypersetup{citecolor=blue}
\newcommand{\dif}{d}

\newcommand{\Eqref}[1]{(\ref{#1})}
\newcommand{\half}{\frac{1}{2}}

\newcommand{\brac}[1]{\left(#1 \right)}
\newcommand{\sbrac}[1]{\left[#1\right]}

\newcommand{\Ecal}{\mathcal{E}}
\newcommand{\Lcal}{\mathcal{L}}
\newcommand{\Ucal}{\mathcal{U}}

\begin{document}

\title{Dynamics of a charged particle around a
weakly magnetized naked singularity}
\author{Gulmina Zaman Babar\footnote{gulmina.zaman@sns.nust.edu.pk}\;\;and Mubasher
Jamil\footnote{mjamil@sns.nust.edu.pk}\\
        \small{\textit{Department of Mathematics,}}\\
        \small{\textit{School of Natural Sciences,}}\\
        \small{\textit{National University of Sciences and Technology}}\\
        \small{\textit{Sector H-12, Islamabad, Pakistan}\vspace{24pt}}\\
        Yen-Kheng Lim\footnote{phylyk@nus.edu.sg}\\
    \small{\textit{Department of Physics, National University of Singapore,}}\\
    \small{\textit{Singapore 117551, Singapore}}}
\date{\today}
\maketitle

\begin{abstract}
  We examine the motion of a charged particle  in the
vicinity of a weakly magnetized naked singularity. The escape velocity
and energy of the particle moving around the naked singularity after
being kicked by another particle or photon are investigated. Also at
 innermost stable circular orbit (ISCO) escape velocity and energy
are examined. Effective potential and angular momentum of the
particle are also discussed. We discuss the center of mass energy
after collision between two particles having same mass and opposite
charges moving along the same circular orbit in the opposite
direction. It is investigated that under what conditions  maximum
energy can be produced as a result of collision.
\end{abstract}

\newpage

\section{Introduction}\label{introduction}
It is believed that particles in the surroundings of black holes are significantly influenced by the strong gravitational pull. Although, other forces that are usually supposed  to be weaker are also at
work near these objects. These include forces  applied by the magnetic field and pressure of the infalling hot gases. Recent investigations have brought up the evidence that magnetic forces can be as powerful as gravity near supermassive black holes. But most of the work till now has been investigated for weak magnetic forces around black holes. The phenomenon of jet formation in black holes is widely under observation nowadays. Adequate amount of energy to establish and strengthen jets is yielded by the matter in accretion disk and rotation of the black hole. It is most likely believed that magnetic fields are responsible for the transfer of energy to the jets \cite{McKinney:2006dy,Narayan:2013qaa, international2009cosmic}.
\par
There are several papers that examine the dynamics of particles
moving around weakly magnetized black holes. Motion of a charged
particle near weakly magnetized Schwarzschild black hole is analyzed
in \cite{Frolov:2010mi,Frolov:2011ea,Zahrani:2013up}. The chaotic motion of a charged particle around
Kerr black hole perturbed by magnetic field is investigated in \cite{Nakamura:2010,Takahashi:2008zh,Kopacek:2010yr,Kopacek:2010at,Preti:2010zza,Frolov:2010cr,Igata:2010ny}. Circular motion of charged
particles around Reissner-Nordstr\"om spacetime is discussed in
\cite{Pugliese:2011py}. Chaotic particle motion in the Majumdar-Papapertou metric
is discussed in \cite{Yurtsever:1994yb,Dettmann:1994dj}. Furthermore, collision between
particles and their escape energies after collision around Kiselev
black hole \cite{Jamil:2014rsa} and a slowly rotating Kerr black hole
\cite{Hussain:2014cba} have been studied.
\par
There are compelling arguments that black hole candidates in astrophysics could well be naked singularities instead of black holes, and that the distinction between the two possiblities may have observational consequences \cite{Virbhadra:1998dy,Virbhadra:2002ju,Virbhadra:2007kw}.
In this paper, we discuss motion of a charged particle around weakly
magnetized naked singularity sourced by a massless scalar field. It is an extension of the
Schwarzschild geometry when a massless scalar field is added to it which deforms the event horizon into a naked singularity. This solution was first discovered by Fisher \cite{Fisher:1948yn}, and subsequently rediscovered by Janis, Newman and Winicour \cite{Janis:1968zz} and others \cite{Wyman:1981bd,Agnese:1985xj,Roberts:1989sk}. Subsequent works eventually show that some of these independently-found solutions are equivalent (see, e.g., Ref.~\cite{Virbhadra:1997ie}). This solution is known in literature as the Fisher solution, or the Janis-Newman-Winicour solution. The understanding of the physical properties and other features of this spacetime were expanded by Janis, Robinson and Winicour, \cite{Janis:1970kn}; therefore it is also sometimes known as the Fisher-Janis-Robinson-Winicour spacetime. A generalization to higher dimensions was given in \cite{Xanthopoulos:1989kb}, and an analysis of its properties in higher dimensions was performed by Abdolrahimi and Shoom \cite{Abdolrahimi:2009dc}. More recently it was shown that this spacetime is isometric to a class of unique spacetimes possessing a photon sphere \cite{Yazadjiev:2015hda}.
\par
While this spacetime is known under many different names, for the purposes of this paper, we shall follow references related to the context of our paper such as \cite{Virbhadra:2002ju,Chowdhury:2011aa, Patil:2011aa} and refer to this solution as the Janis-Newman-Winicour (JNW) spacetime. The charged generalization of this solution was obtained by Penney \cite{Penney:1968zz}, and an exact magnetized solution was given by \cite{Astorino:2013xc}, and a further accelerating generalization was given in \cite{Astorino:2013sfa}.
\par
In \cite{Zahrani:2013up} motion of a charged particle is discussed in the vicinity of weakly magnetized Schwarzschild black hole. The effect of the magnetic field on a charged particle is similar to the effect of the black hole rotation on a neutral particle. Like fast-rotating black holes weakly magnetized black holes behaves as particle accelerators. The case of particle motion around weakly magnetized Kerr black holes was considered recently in \cite{Shiose:2014bqa}. The critical escape energy and velocity of the kicked charged particle with different initial radial velocities have been investigated in \cite{Huang:2014tra}. Recently Lim studied the motion of charged particles around an electrified black hole \cite{Lim:2015oha}.
\par
Our core interest is to study the motion of a charged particle
orbiting in the JNW spacetime after being hit by another particle.
We are considering homogeneous magnetic field at spatial infinity.
After collision the particle may pursue its motion in distinctive
trajectories. We examine that,  under what circumstances can a
particle leave the locality of naked singularity and what causes the
particle to fall in the naked singularity.

In Sec.~\ref{mjnw}, we derive the general equations of motion for a charged particle in a magnetized JNW spacetime. The equations will contain special cases that will be studied in subsequent sections. Particularly, in Sec.~\ref{jnw} we consider the special case of a particle
in the absence of a magnetic field. The value of inner most stable
circular  orbit (ISCO) is investigated. Escape velocity of the
particle is determined, also at the ISCO its value is computed.
Energy of the particle is brought under analysis and its value at
the ISCO is calculated. In Sec.~\ref{wmjnw}, the motion of charged particles in
weakly magnetized spacetimes are discussed. To simplify the analysis, dimensionless forms of the equations of motion,
effective potential, energy and escape velocity are discussed.
Dimensionless angular momentum and magnetic field are also
investigated. In Sec.~\ref{collision}, we made an analysis of collision
between two particles with  same mass and opposite charges revolving
around the naked singularity and computed the center-of-mass energy
produced as a result of  collision.

\section{Particles in a magnetized JNW spacetime} \label{mjnw}

\subsection{Metric}

We will be considering the motion of a charged particle in the magnetized JNW (or FJRW) spacetime \cite{Astorino:2013xc}. This can be obtained by taking the ordinary JNW metric and applying a Harrison-type transformation \cite{Harrison:1968,Dowker:1993bt,Ortaggio:2004kr,Astorino:2013xc} to it. The result is an exact solution given by the metric
\begin{align}
 \dif s^2&=\Lambda^2\brac{-f^\nu\dif t^2+f^{-\nu}\dif r^2+r^2f^{1-\nu}\dif\theta^2}+\Lambda^{-2}r^2f^{1-\nu}\sin^2\theta\,\dif\phi^2,\nonumber\\
 f&=1-\frac{r_g}{r},\quad \Lambda=1+\frac{1}{4}B^2r^2f^{1-\nu}\sin^2\theta, \label{MJNW_metric}
\end{align}
while the massless scalar field is given by
\begin{align}
 \varphi&=\sqrt{\frac{1-\nu^2}{2}}\ln f,\label{MJNW_scalar}
\end{align}
and the Maxwell potential is
\begin{align}
 A&=\frac{Br^2f^{1-\nu}\sin^2\theta}{2\Lambda}\,\dif\phi.\label{MJNW_Max}
\end{align}
The Maxwell tensor accordingly is $F_{\mu\nu}=\partial_\mu A_\nu-\partial_\nu A_\mu$. It can be verified that the metric \Eqref{MJNW_metric}, together with \Eqref{MJNW_scalar} and \Eqref{MJNW_Max} solves the four-dimensional Einstein-Maxwell-scalar equations.
\par
The parameters $r_g$ and $\nu$ are related to the black hole mass $M$ and scalar charge $q$ by
\begin{align}
 M=\half\nu r_g,\quad q=\half r_g\sqrt{1-\nu^2},
\end{align}
while the parameter $B$ is related to the strength of the axisymmetric magnetic field. We can see that the case $\nu=1$ (or, equivalently, $q=0$) reduces to the magnetic Ernst spacetime. On the other hand, for $B=0$ we recover the usual JNW spacetime. Setting both $B=0$ and $\nu=1$ gives the Schwarzschild spacetime. Similar to the unmagnetized case, the metric \Eqref{MJNW_metric} has a strong curvature singularity at $r=r_g$. Similar to the Ernst solution, this solution is not asymptotically flat.

\subsection{Equations of motion} \label{mjnw_eom}
For a test particle of charge per unit mass $e$, its motion is described by a trajectory $x^\mu(\tau)$, where $\tau$ is an appropriate affine parametrization. The motion is determined by the Lagrangian $L=\half g_{\mu\nu}\dot{x}^\mu\dot{x}^\nu+eA_\mu\dot{x}^\mu$, where overdots denote derivatives with respect to $\tau$. The equations of motion satisfied by $x^\mu(\tau)$ can be derived with the Euler-Lagrange equations $\frac{\dif}{\dif\tau}\frac{\partial L}{\partial\dot{x}^\mu}=\frac{\partial L}{\partial x^\mu}$.
\par
In the following, it will be convenient to define
\begin{align}
 F\equiv f^\nu,\quad G\equiv f^{1-\nu},
\end{align}
such that the Lagrangian describing a charged particle in the magnetized JNW spacetime is given by
\begin{align}
 L=&\;\half\sbrac{\Lambda^2\brac{-F\dot{t}^2+\frac{\dot{r}^2}{F}+r^2G\dot{\theta}^2}+\frac{r^2G\sin^2\theta}{\Lambda^2}\dot{\phi}^2}+\frac{eBr^2G\sin^2\theta}{2\Lambda}\dot{\phi}.
\end{align}
Since the Lagrangian is cyclic in $t$ and $\phi$, we have the following first integrals
\begin{align}
 \dot{t}=\frac{\Ecal}{\Lambda^2F},\quad\dot{\phi}=\frac{\Lambda^2}{r^2G\sin^2\theta}\brac{\Lcal-\frac{eBr^2G\sin^2\theta}{2\Lambda}}, \label{tdotphidot}
\end{align}
where $\Ecal$ and $\Lcal$ are conserved quantities which we may interpret as the energy and angular momentum of the particle, respectively. Applying the Euler-Lagrange equation to the remaining two coordinates gives
\begin{align}
 \ddot{r}&=\brac{\frac{F'}{2F}-\frac{\partial_r\Lambda}{\Lambda}}\dot{r}^2+FG\brac{r+\frac{r^2\partial_r\Lambda}{\Lambda}+\frac{r^2G'}{2G}}\dot{\theta}^2-\frac{2\partial_\theta\Lambda}{\Lambda}\dot{r}\dot{\theta}\nonumber\\
            &\quad+\frac{F}{r^3G\sin^2\theta}\brac{1-\frac{r\partial_r\Lambda}{\Lambda}+\frac{rG'}{2G}}\brac{\Lcal-\frac{eBr^2G\sin^2\theta}{2\Lambda}}^2\nonumber\\
            &\quad+\frac{eBF}{r\Lambda}\brac{1-\frac{r\partial_r\Lambda}{2\Lambda}+\frac{rG'}{2G}}\brac{\Lcal-\frac{eBr^2G\sin^2\theta}{2\Lambda}}-\brac{\frac{\partial_r\Lambda}{\Lambda}+\frac{F'}{2F}}\frac{\Ecal^2}{\Lambda^4},\label{rddot}\\
  \ddot{\theta}&=\frac{\partial_\theta\Lambda}{\Lambda}\brac{\frac{\dot{r}^2}{Fr^2G}-\dot{\theta}^2}-2\brac{\frac{\partial_r\Lambda}{\Lambda}+\frac{1}{r}+\frac{G'}{2G}}\dot{r}\dot{\theta}-\frac{\Ecal^2\partial_\theta\Lambda}{\Lambda^5Fr^2G}\nonumber\\
     &\quad+\frac{1}{r^4G^2\sin^3\theta}\brac{\cos\theta-\frac{\sin\theta\partial_\theta\Lambda}{\Lambda}}\brac{\Lcal-\frac{eBr^2G\sin^2\theta}{2\Lambda}}^2\nonumber\\
     &\quad+\frac{eB}{\Lambda r^2G\sin\theta}\brac{\cos\theta-\frac{\sin\theta\partial_\theta\Lambda}{2\Lambda}}\brac{\Lcal-\frac{eBr^2G\sin^2\theta}{2\Lambda}}. \label{thetaddot}
\end{align}
Here, the primes appearing in $F'$ and $G'$ denote derivatives with respect to $r$. The invariance of $g_{\mu\nu}\dot{x}^\mu\dot{x}^\nu\equiv\epsilon$ gives a constraint equation
\begin{align}
 -\frac{\Ecal^2}{\Lambda^2F}+\frac{\Lambda^2}{r^2G\sin^2\theta}\brac{\Lcal-\frac{eBr^2G\sin^2\theta}{2\Lambda}}^2+\Lambda^2\brac{\frac{\dot{r}^2}{F}+r^2G\dot{\theta}^2}=\epsilon. \label{first_integral}
\end{align}
By appropriately rescaling the affine parameter $\tau$, the magnitude of $\epsilon$ can be set to unity if it is nonzero. Therefore in the following, we have $\epsilon=-1$ for timelike particles and $\epsilon=0$ for photons or null geodesics. The equations of motion \Eqref{rddot} and \Eqref{thetaddot}, together with the constraint \Eqref{first_integral} contains various special cases for motion in related spacetimes.  For instance, setting $\nu=1$ describes the motion in the magnetic Ernst spacetime studied in Refs.~\cite{Dadhich1979,Estaban1984,Karas1990,Lim:2015oha}. Setting $B=0$ will give the geodesic equations in the JNW spacetime, and finally if $\nu=1$ and $B=0$ we have the well-known equations for Schwarzschild geodesics.
\par
The constraint equation \Eqref{first_integral} can be also cast in the effective potential formulation
\begin{align}
 &\Lambda^4\brac{\dot{r}^2+r^2FG\dot{\theta}^2}=\Ecal^2-\mathcal{U}^2,\label{COE}
\end{align}
where
\begin{align}
 \mathcal{U}^2=\frac{\Lambda^4F}{r^2G\sin^2\theta}\brac{\Lcal-\frac{eBr^2G\sin^2\theta}{2\Lambda}}^2-\epsilon\Lambda^2F. \label{U2_eff}
\end{align}
While in general, the effective potential $\Ucal^2$ is a complicated function that depends on parameters $\Lcal$, $\nu$, $B$ and $r_g$, we can make a few qualitative observations.
\par
Firstly, we note the asymptotic behavior of $\Ucal^2$, at large $r$, is
\begin{align}
 \Ucal^2\sim\frac{1}{256}B^6\brac{r\sin\theta}^6\brac{\Lcal B-2e}^2+\mathcal{O}(r^5).
\end{align}
Thus, at large distances away from the singularity, the potential increases to the order of $\brac{r\sin\theta}^6$. Therefore, for large values of $r$ where $r\rightarrow\infty$, the potential remains finite if $\theta\rightarrow 0$ or $\pi$. Thus a particle with finite energy $\Ecal$ can escape to infinity if the trajectory takes it along a path close to the polar axis $\theta=0$ or $\theta=\pi$.
\par
Secondly, at distances close to the $r=r_g$ singularity, due to the factor $F/G=f^{2\nu-1}$ in front of the paranthesis in \Eqref{U2_eff}, we have
\begin{align}
 \lim_{r\rightarrow r_g^+}\Ucal^2=\left\{\begin{array}{cl}
                                         0, & \mbox{if } \nu>\half, \\
                                         +\infty & \mbox{if }\nu<\half \mbox{ and } \brac{\Lcal-\frac{eBr^2G\sin^2\theta}{2\Lambda}}\neq 0.
                                         \end{array}\right. \label{inf_barrier}
\end{align}
Therefore for $\nu<\half$ and $\brac{\Lcal-\frac{eBr^2G\sin^2\theta}{2\Lambda}}\neq 0$ there exist an infinite potential barrier preventing the particle from reaching the $r=r_g$ singularity.

\subsection{Equatorial circular orbits}
\begin{figure}
 \begin{center}
  \begin{subfigure}[b]{0.31\textwidth}
   \centering
   \includegraphics[scale=0.25]{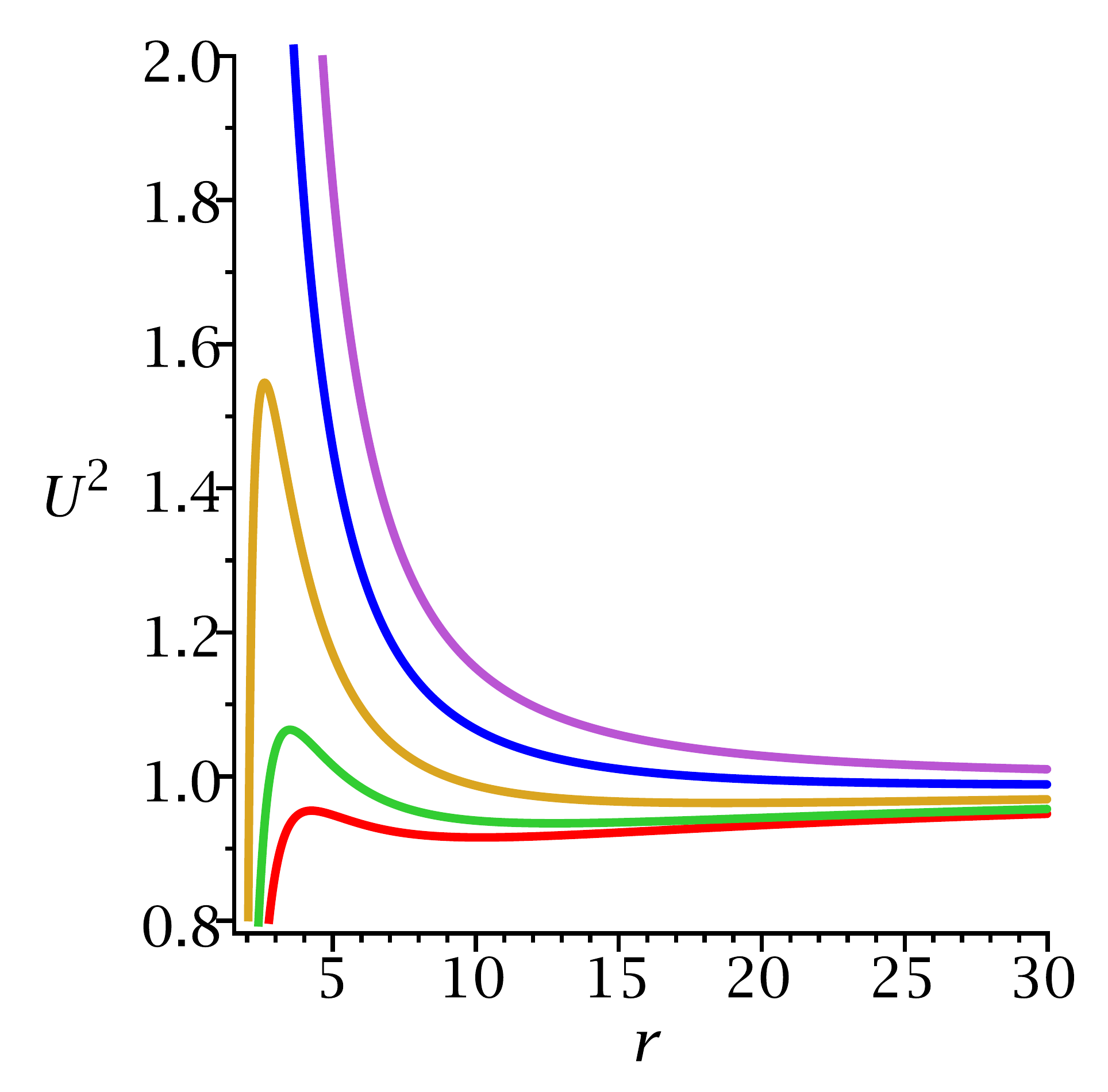}
   \caption{$B=0$, $\Lcal=3.75$.}
   \label{Ueq-11}
  \end{subfigure}
  \begin{subfigure}[b]{0.31\textwidth}
   \centering
   \includegraphics[scale=0.25]{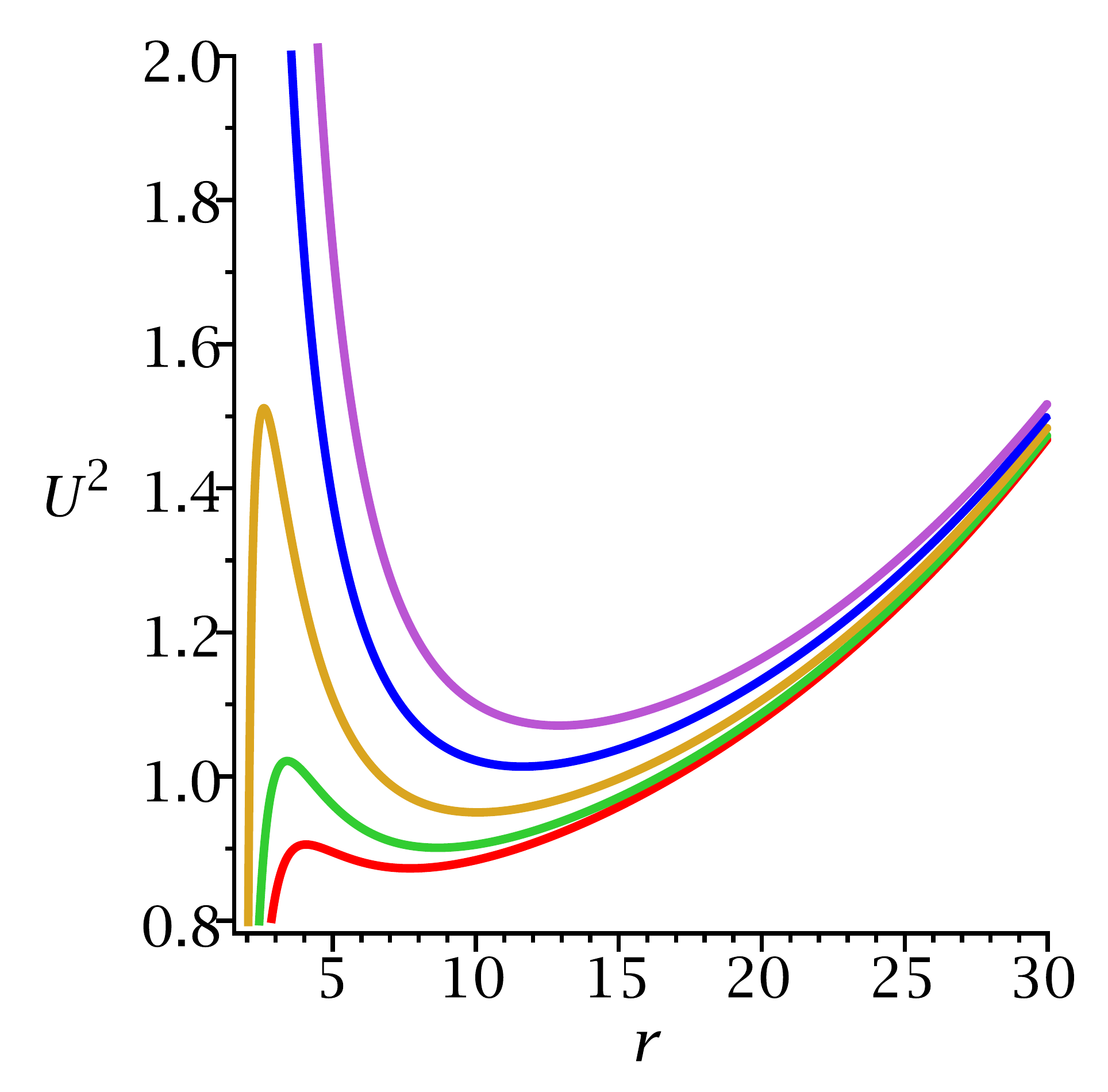}
   \caption{$B=0.03$, $\Lcal=3.75$.}
   \label{Ueq-12}
  \end{subfigure}
  \begin{subfigure}[b]{0.31\textwidth}
   \centering
   \includegraphics[scale=0.25]{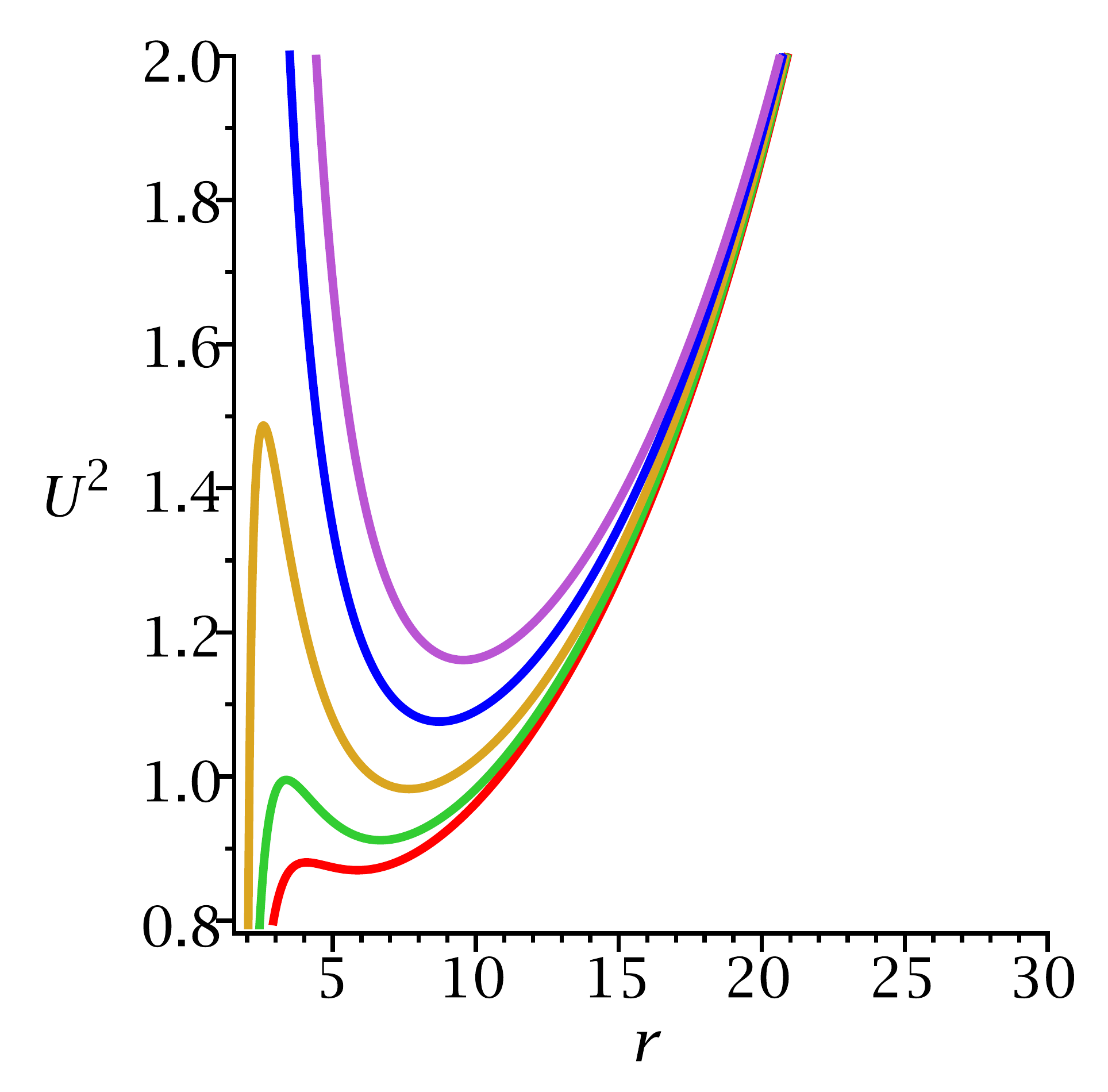}
   \caption{$B=0.06$, $\Lcal=3.75$.}
   \label{Ueq-13}
  \end{subfigure}
  \begin{subfigure}[b]{0.31\textwidth}
   \centering
   \includegraphics[scale=0.25]{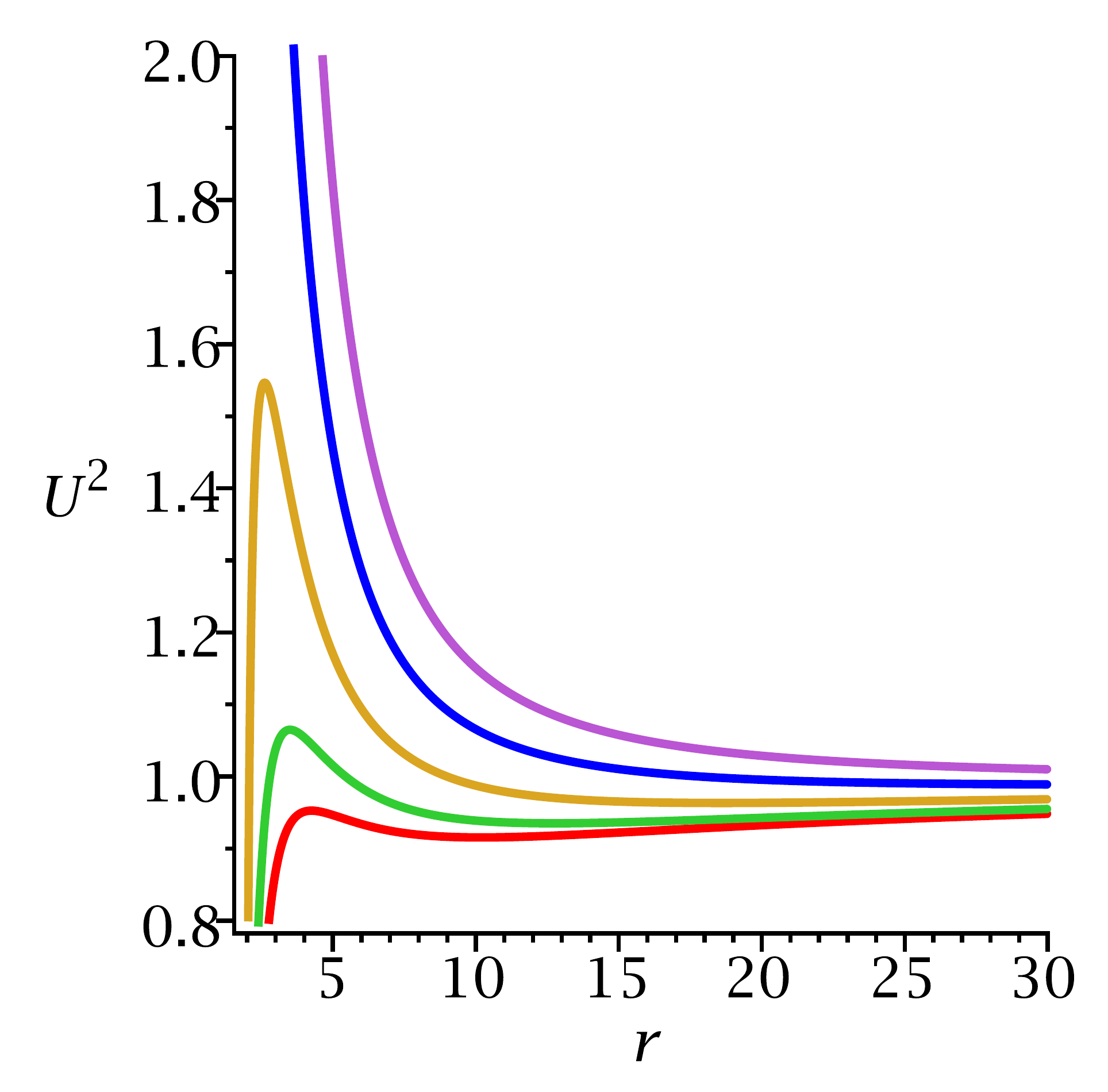}
   \caption{$B=0$, $\Lcal=-3.75$.}
   \label{Ueq-21}
  \end{subfigure}
  \begin{subfigure}[b]{0.31\textwidth}
   \centering
   \includegraphics[scale=0.25]{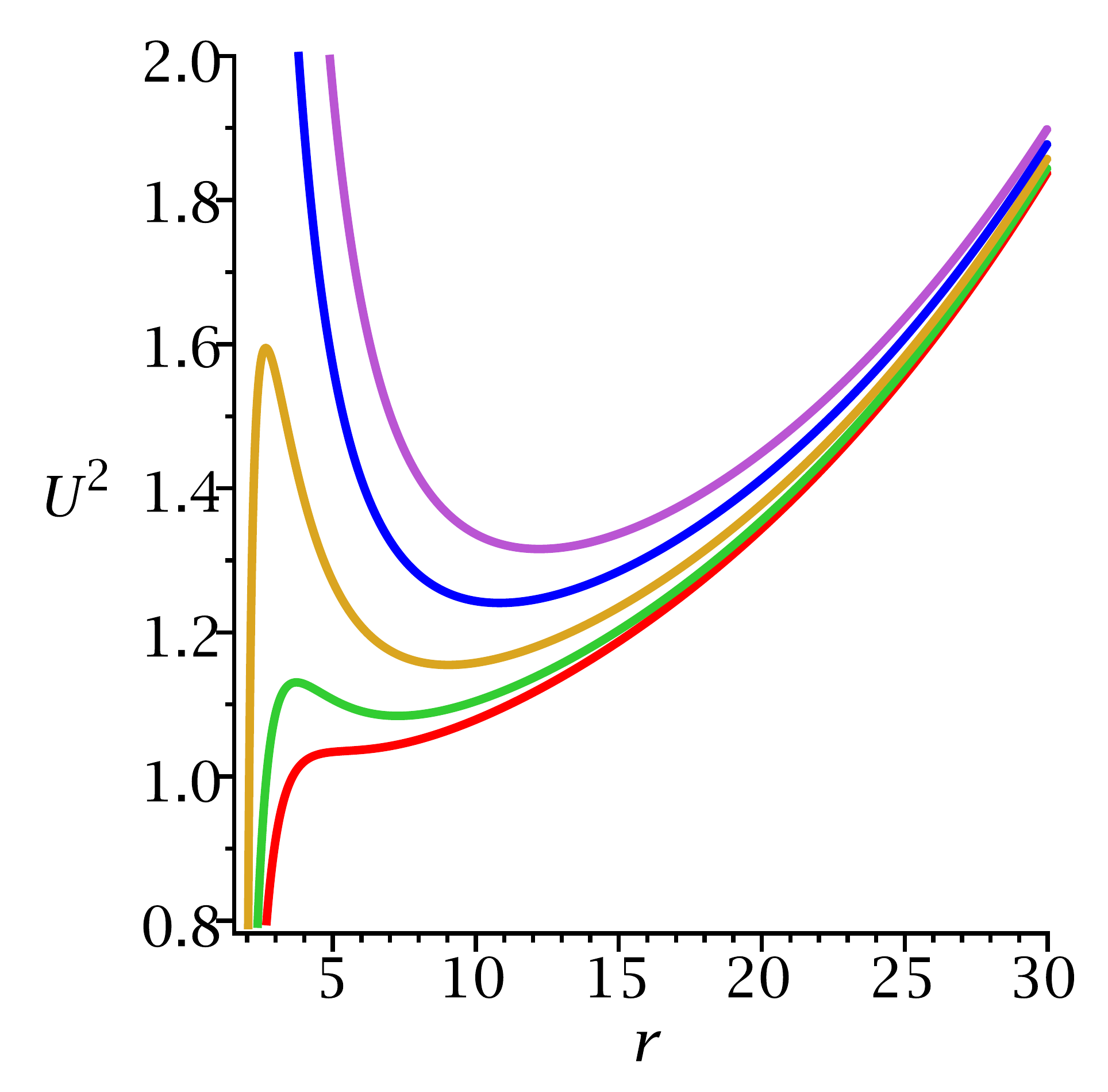}
   \caption{$B=0.03$, $\Lcal=-3.75$.}
   \label{Ueq-22}
  \end{subfigure}
  \begin{subfigure}[b]{0.31\textwidth}
   \centering
   \includegraphics[scale=0.25]{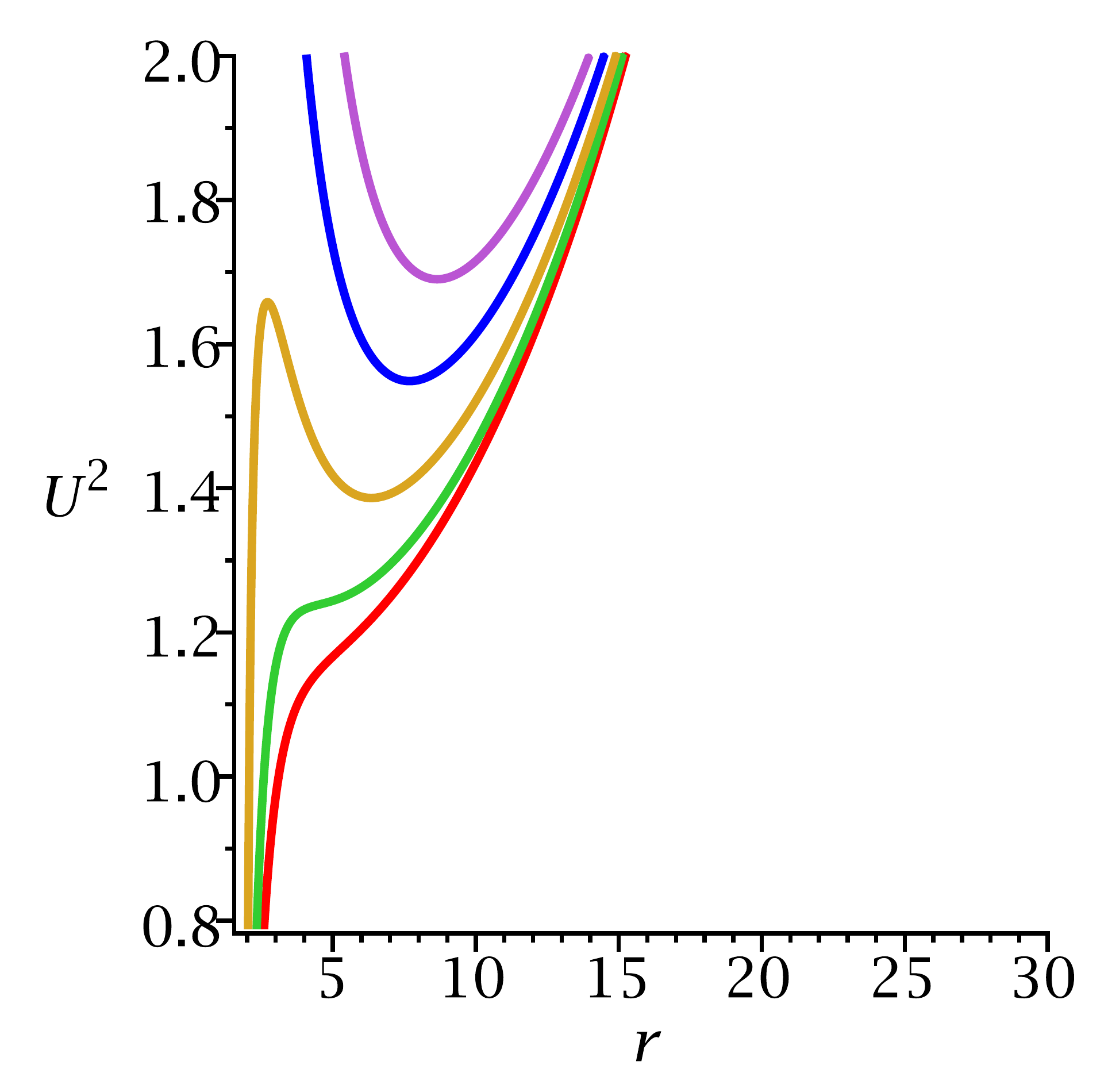}
   \caption{$B=0.06$, $\Lcal=-3.75$}
   \label{Ueq-23}
  \end{subfigure}
  \caption{(Color online) Plots of $\mathcal{U}^2$ vs $r$ at $\theta=\pi/2$ and $r_g=2$ for timelike ($\epsilon=-1$) particles of charge $e=1$, with various values of $B$ and $\Lcal$. Each plot contains curves of various values of $\nu$, from bottom to the top, $\nu=$ 1, 0.9, 0.7, 0.5 and 0.3.}
  \label{Ueq}
 \end{center}
\end{figure}

We can easily check that $\theta=\pi/2$ is a trivial solution to
Eq.~\Eqref{thetaddot}, where $\ddot{\theta}=\dot{\theta}=0$. These
orbits lie in the equatorial plane. The plots of $\mathcal{U}^2$
against $r$ for various cases are shown in Fig. (\ref{Ueq}).
\par
For the unmagnetized case $B=0$, the effective potential obtained in \cite{Chowdhury:2011aa} is reproduced. In particular, for $\nu<\frac{1}{2}$ the potential becomes an infinite barrier. Now if the magnetic field is turned on, an additional gravitational attraction is provided by the magnetic field itself, manifesting in a potential well. Thus charged particles in the magnetized JNW spacetime is bound more tightly compared to the unmagnetized case.

For concreteness, we will be mainly  interested in a charged
particles which are initially in the innermost stable circular orbit
(ISCO), while second particle is coming from infinity. They both
collide in the ISCO. We may define circular orbits to be those of
constant $r$ and $\theta$. We seek such orbits exist in the
equatorial plane where $\theta$ is a constant equal to $\pi/2$, thus
Eq.~\Eqref{thetaddot} is automatically satisfied. By demanding that
$r=r_o$ is a constant in \Eqref{rddot} and \Eqref{first_integral},
we have,
\begin{align}
 \Lcal_\pm&=\frac{2r_oG\sbrac{eBr_o^2G\brac{6B^2r_oFG+3B^2r_o^2FG'+F'(4+B^2r_o^2G)}\pm\sqrt{K}}}{(4+B^2r^2G)\sbrac{(3B^2r_o^2G-4)(2G+r_oG')F+r_oGF'(4+B^2r_o^2G)}}, \label{L_circ}\\
   \Ecal^2_\pm&=\frac{F(4+B^2r_o^2G)^2}{256r_o^2G}\nonumber\\
               &\quad\times\sbrac{(4+B^2r_o^2G)^2\Lcal_\pm^2-4eBGr_o^2(4+B^2r_o^2G)\Lcal_\pm+4Gr_o^2(e^2B^2r_o^2G-4\epsilon)}, \label{E_circ}
\end{align}
where
\begin{align}
 K=&\;2B^2r_o^2(2G+r_oG')^2\brac{3r_o^2G\epsilon B^2-4\epsilon+2e^2}F^2\nonumber\\
     &\;+\epsilon r_oF'\brac{4+B^2r_o^2G}\brac{5B^2r_o^2G-4}\brac{2G+r_oG'}F+r_o^2GF'^2\epsilon\brac{4+B^2r_o^2G}^2
\end{align}
We consider the stability of a circular orbit  by perturbing about
the radius $r_o$ by writing
\begin{align}
 r(\tau)=r_o+\varepsilon r_1(\tau).
\end{align}
We substitute this into Eq.~\Eqref{rddot}. To first order in $\varepsilon$, the equations of motion reduce to
\begin{align}
 \ddot{r}_1=&\;-\omega^2 r_1, \label{circ_perturb}
\end{align}
where, in general, the expression for $\omega$ is a complicated function of $\nu$, $B$, $r_g$ and $r_0$ that is too complicated to be shown here, however it can be handled straightforwardly using a standard computer algebra software such as MAPLE.
\par
\begin{figure}
 \begin{center}
  \begin{subfigure}[b]{0.31\textwidth}
   \centering
   \includegraphics[scale=0.25]{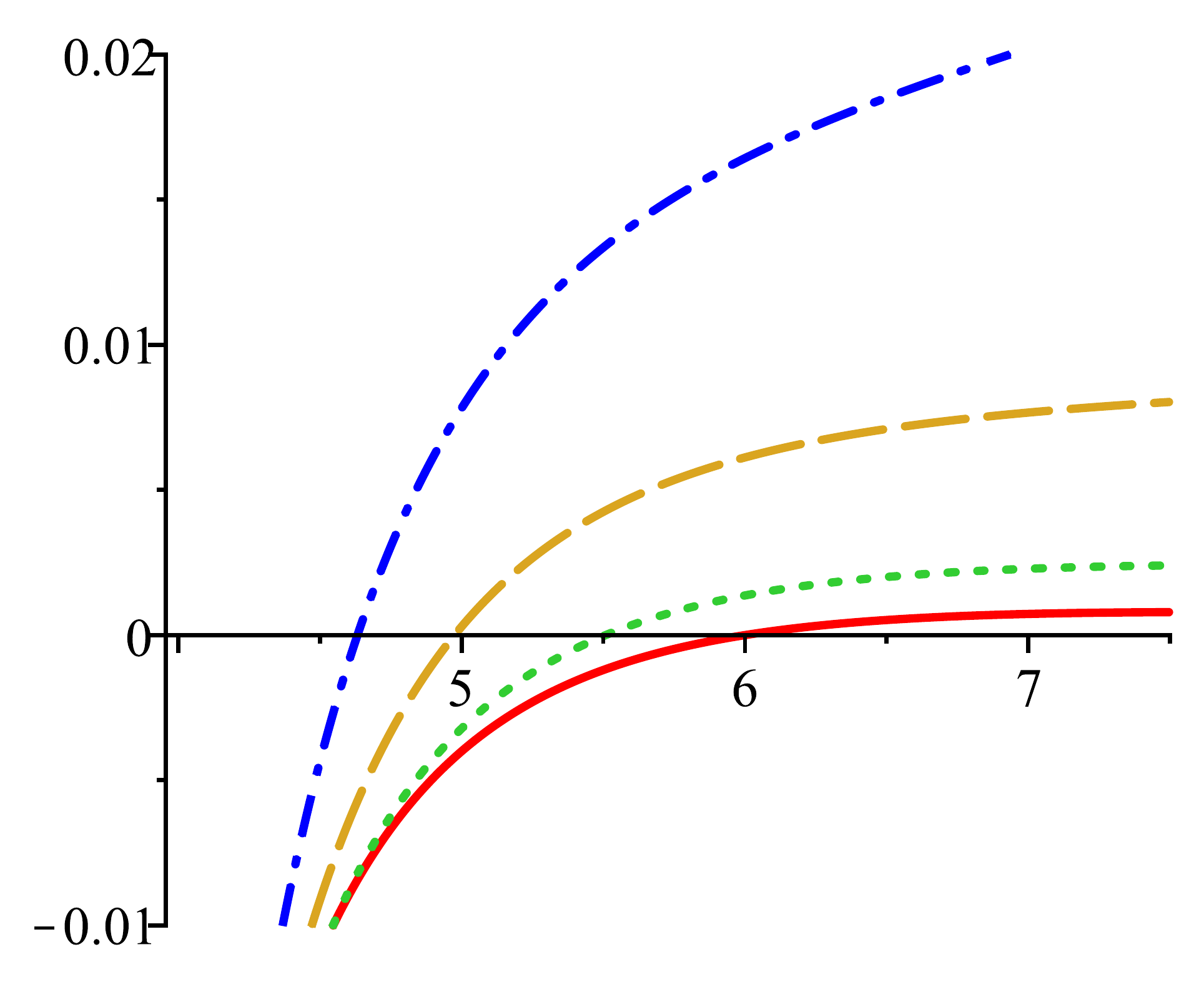}
   \caption{$\nu=1$, $\Lcal=\Lcal_+$.}
   \label{omega2-11}
  \end{subfigure}
  \begin{subfigure}[b]{0.31\textwidth}
   \centering
   \includegraphics[scale=0.25]{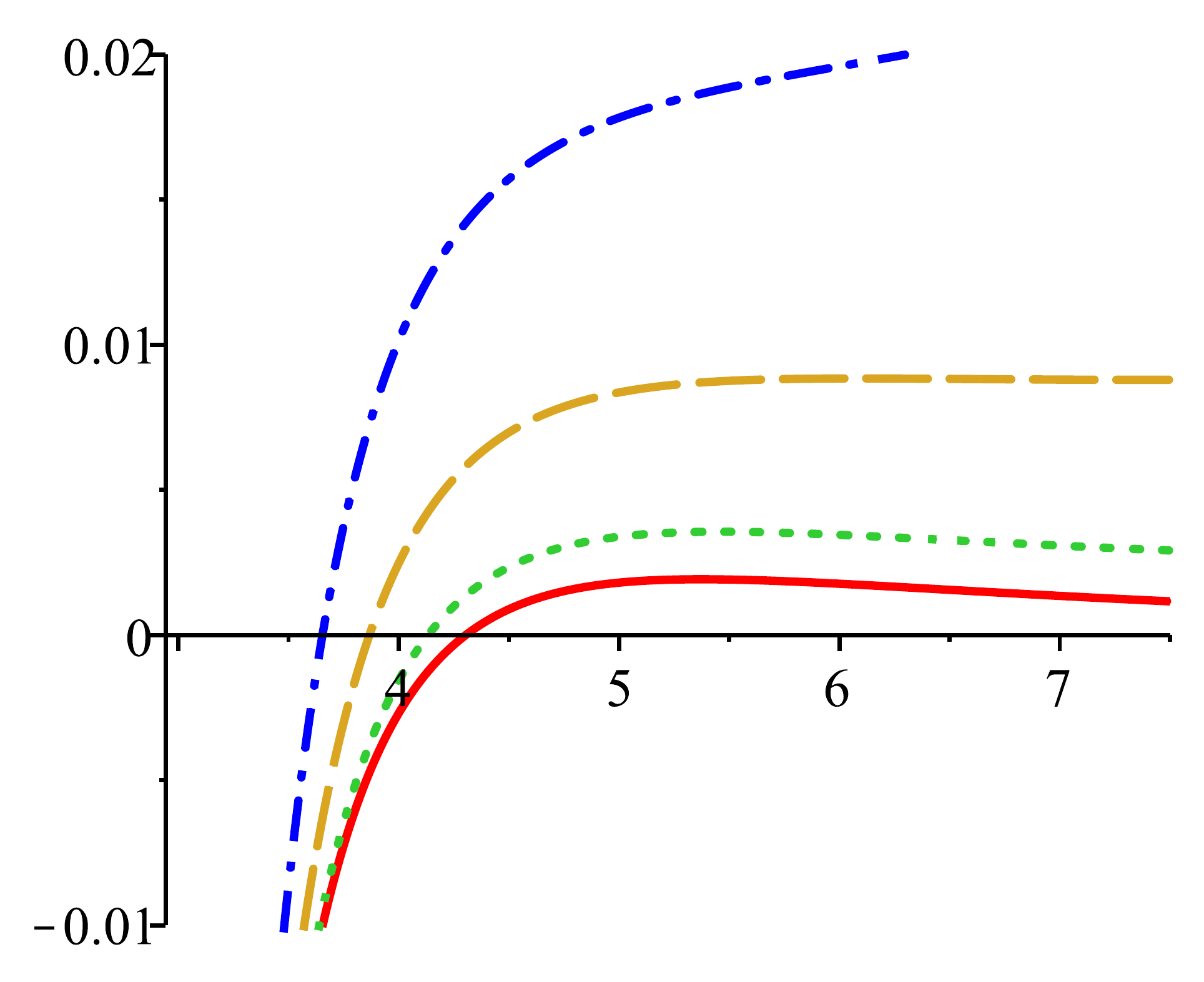}
   \caption{$\nu=0.7$, $\Lcal=\Lcal_+$.}
   \label{omega2-12}
  \end{subfigure}
  \begin{subfigure}[b]{0.31\textwidth}
   \centering
   \includegraphics[scale=0.25]{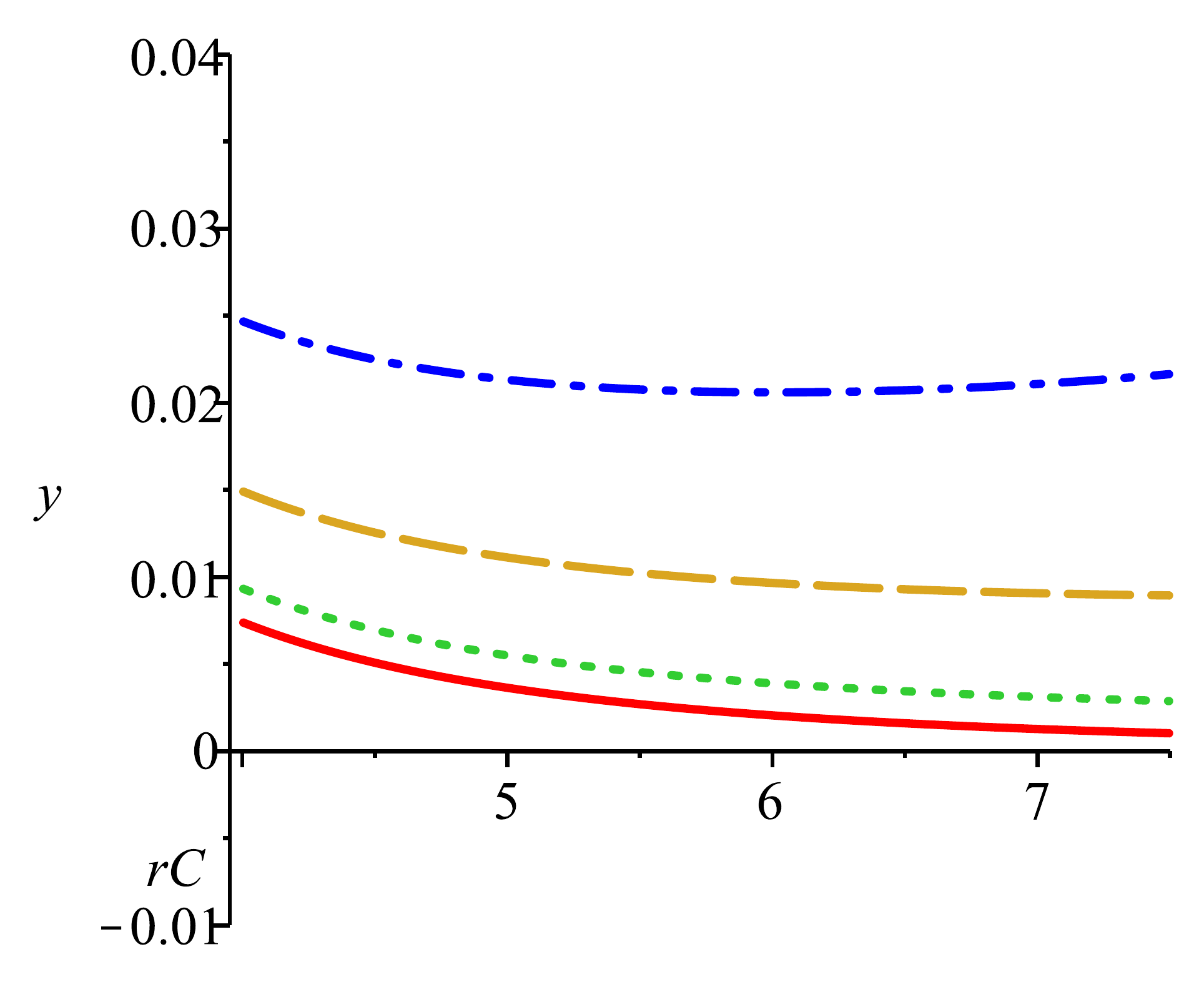}
   \caption{$\nu=0.4$, $\Lcal=\Lcal_+$.}
   \label{omega2-13}
  \end{subfigure}
  \begin{subfigure}[b]{0.31\textwidth}
   \centering
   \includegraphics[scale=0.25]{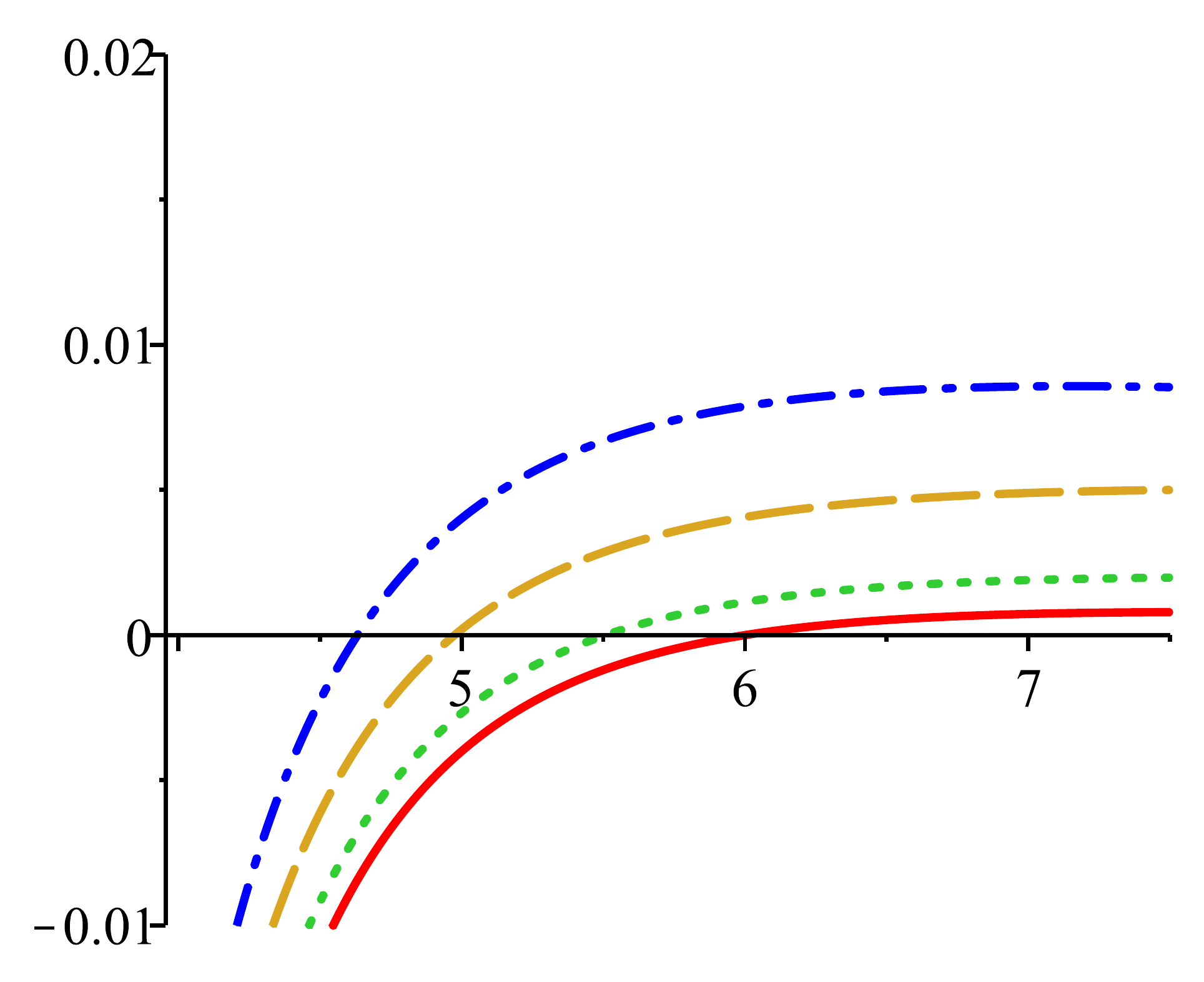}
   \caption{$\nu=1$, $\Lcal=\Lcal_-$.}
   \label{omega2-21}
  \end{subfigure}
  \begin{subfigure}[b]{0.31\textwidth}
   \centering
   \includegraphics[scale=0.25]{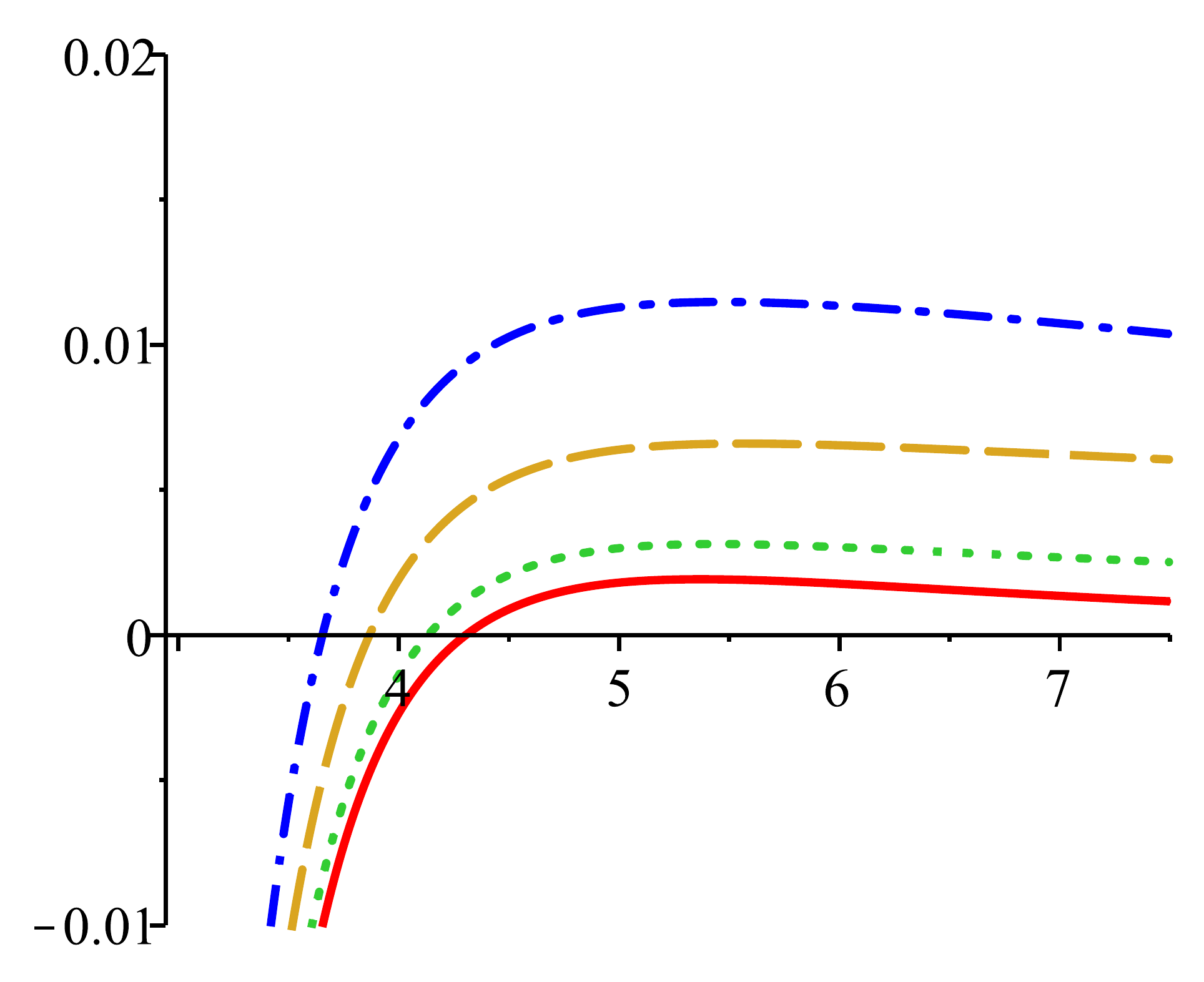}
   \caption{$\nu=0.7$, $\Lcal=\Lcal_-$.}
   \label{omega2-22}
  \end{subfigure}
  \begin{subfigure}[b]{0.31\textwidth}
   \centering
   \includegraphics[scale=0.25]{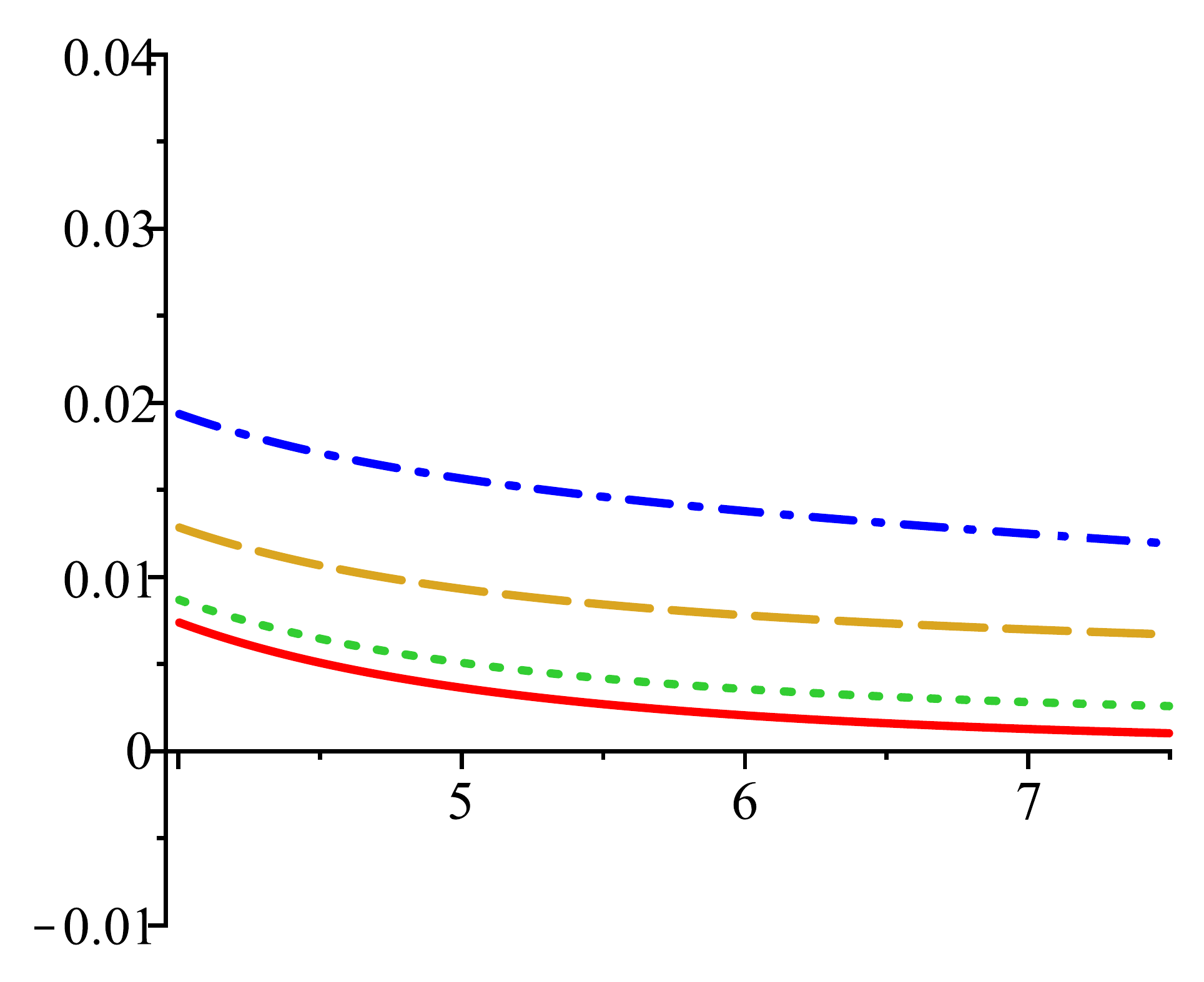}
   \caption{$\nu=0.4$,  $\Lcal=\Lcal_-$.}
   \label{omega2-23}
  \end{subfigure}
  \caption{(Color online) Plots of $\omega^2$ vs $r_0$, for $e=1$ and various $\nu$ in units where $r_g=2$. For each value of $\nu$, the solid, dotted, dashed and dash-dotted curves respectively correspond to $B=$ 0, 0.025, 0.05, and 0.75.}
  \label{omega2}
 \end{center}
\end{figure}

We can, however, study the qualitative behavior of $\omega^2$ for
various spacetime and orbital parameters. Fig. (\ref{omega2}) shows
the values of $\omega^2$ for various $B$ and $\nu$. The circular
orbits are stable for $\omega^2>0$, and we can see that for the case
$B=0$ and $\nu=1$ (the solid curve in Fig. (\ref{omega2-11})) show
that the stable circular orbits exist for $r_o>3r_g$. This is the
well-known case of circular orbits in the Schwarzschild spacetime.
\par
Turning on the magnetic field, while keeping $\nu=1$ corresponds to
circular orbits in the magnetic Ernst spacetime. We see that the
effect of the magnetic field brings the stable range of circular
orbits closer to the black hole. (See the dotted, dashed and
dash-dotted curves in Fig. (\ref{omega2-11}) and (\ref{omega2-21})).
\par
Setting $\nu<1$, we see that spacetimes with a massless scalar field
can support stable circular orbits of even smaller radii, in
particular, for $\nu<1/2$ all circular orbits up to those
infinitesimally close to the singularity $r=r_g$ are stable. This is
due to the existence of the infinite potential barrier, as shown in
\Eqref{inf_barrier}.

\subsection{Escape trajectories} \label{esc_trajectories_mjnw}
We consider particles initially in an ISCO which subsequently
collides with another particle. In general, the collision would
alter its energy and angular momentum, as well as its radial and
angular velocities. For concreteness and simplicity, we restrict our
attention to collision which only changes the energy of the
particle, while the angular momentum remains the same. Furthermore,
we assume that the radial velocity remains unchanged, therefore,
$\dot{r}=0$ after collision. From \Eqref{first_integral}, this
restricts us to the fact that increasing the energy results
$\dot{\theta}\neq0$, i.e., the particle gets kicked out from the
equatorial plane.
\par
Plotting the curves of $\Ucal^2=\Ecal^2$ shows the boundary of
regions accessible to the particle after collision. To demonstrate a
specific example, we consider particles kicked from an ISCO in a
spacetime of $B=0.005$, $\nu=0.9$, $r_g=2$ and $e=10$. Choosing the
lower sign in \Eqref{L_circ}, and solving for $\omega^2=0$ using
Eq.\Eqref{circ_perturb}, we find that the initial ISCO radius is
$r_{\mathrm{(ISCO)}}=4.9490145$, and the angular momentum and energy
are respectively $\Lcal_{-(\mathrm{ISCO})}=2.986451488$ and
$\Ecal_{(\mathrm{ISCO})}=0.8853807128$. With these parameters the
curves in $\Ucal^2=\Ecal^2$ are plotted in Cartesian-type
coordinates projected on a plane where $\phi=0$, (or, $y=0$).
\par
\begin{figure}
 \begin{center}
  \begin{subfigure}[b]{0.4\textwidth}
   \centering
   \includegraphics[scale=0.3]{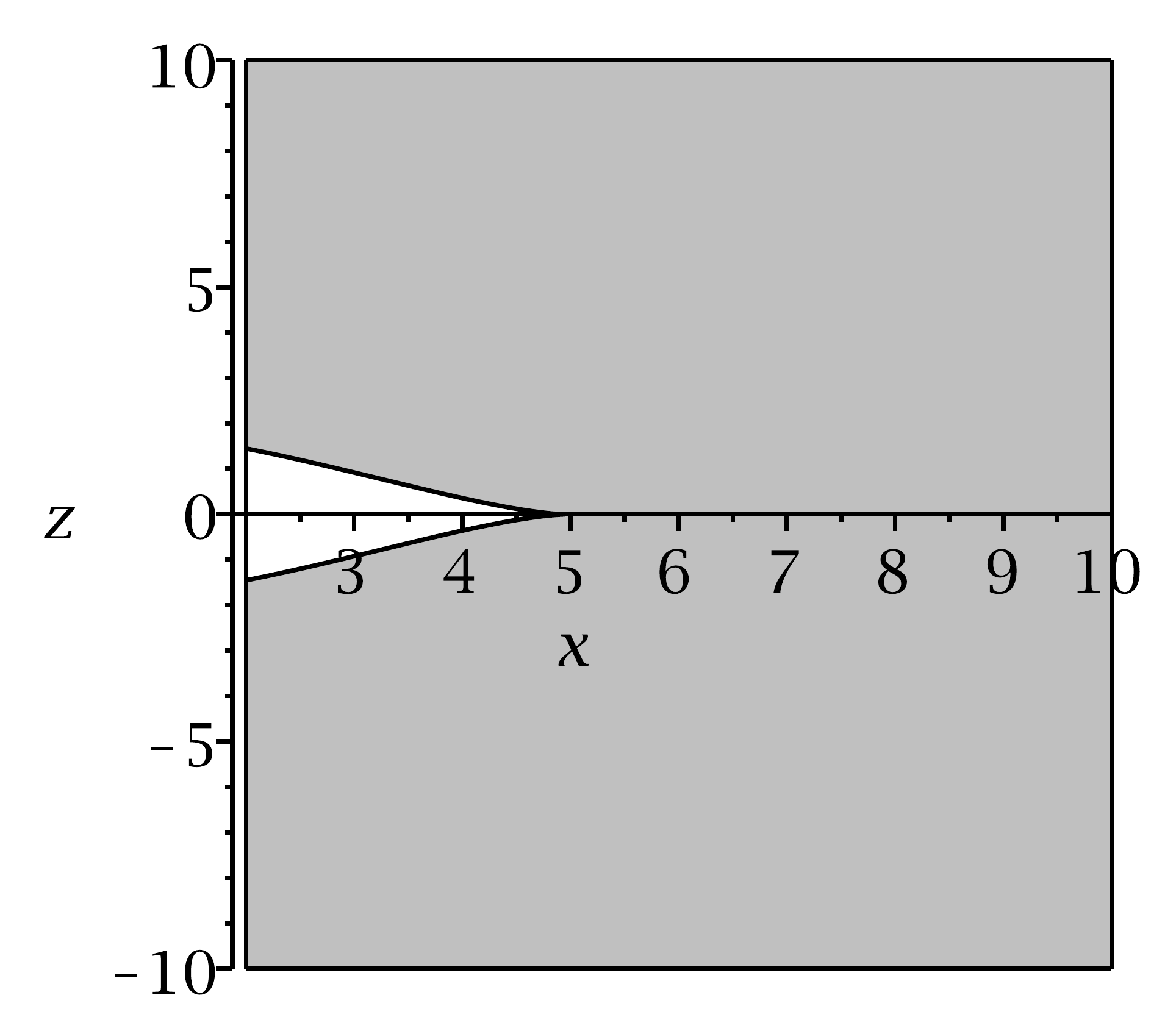}
   \caption{$\Ecal=\Ecal_{(\mathrm{ISCO})}$.}
   \label{esc-1}
  \end{subfigure}
  \begin{subfigure}[b]{0.4\textwidth}
   \centering
   \includegraphics[scale=0.3]{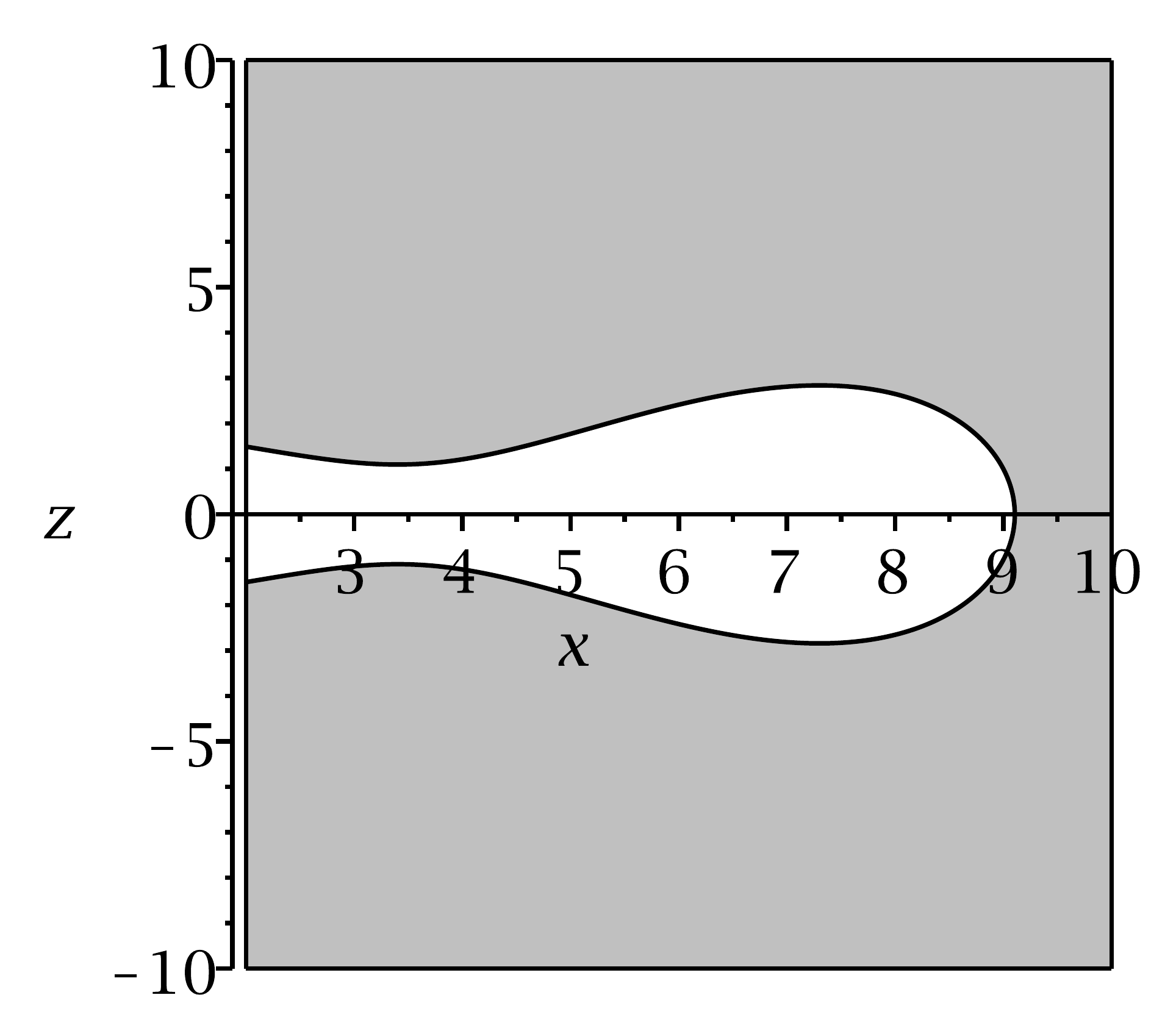}
   \caption{$\Ecal=0.9$.}
   \label{esc-2}
  \end{subfigure}
  \begin{subfigure}[b]{0.4\textwidth}
   \centering
   \includegraphics[scale=0.3]{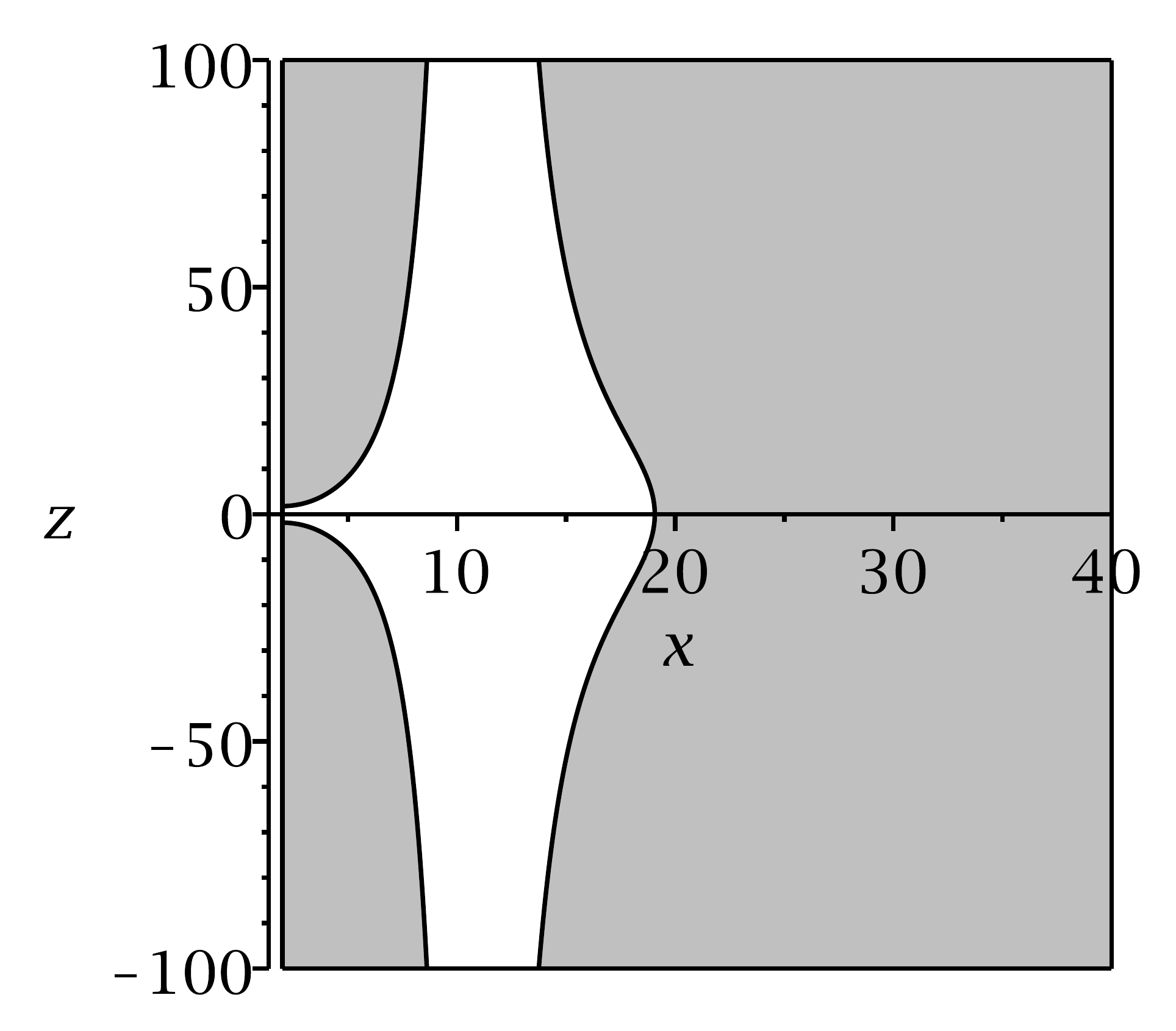}
   \caption{$\Ecal=1.0$.}
   \label{esc-3}
  \end{subfigure}
  \begin{subfigure}[b]{0.4\textwidth}
   \centering
   \includegraphics[scale=0.3]{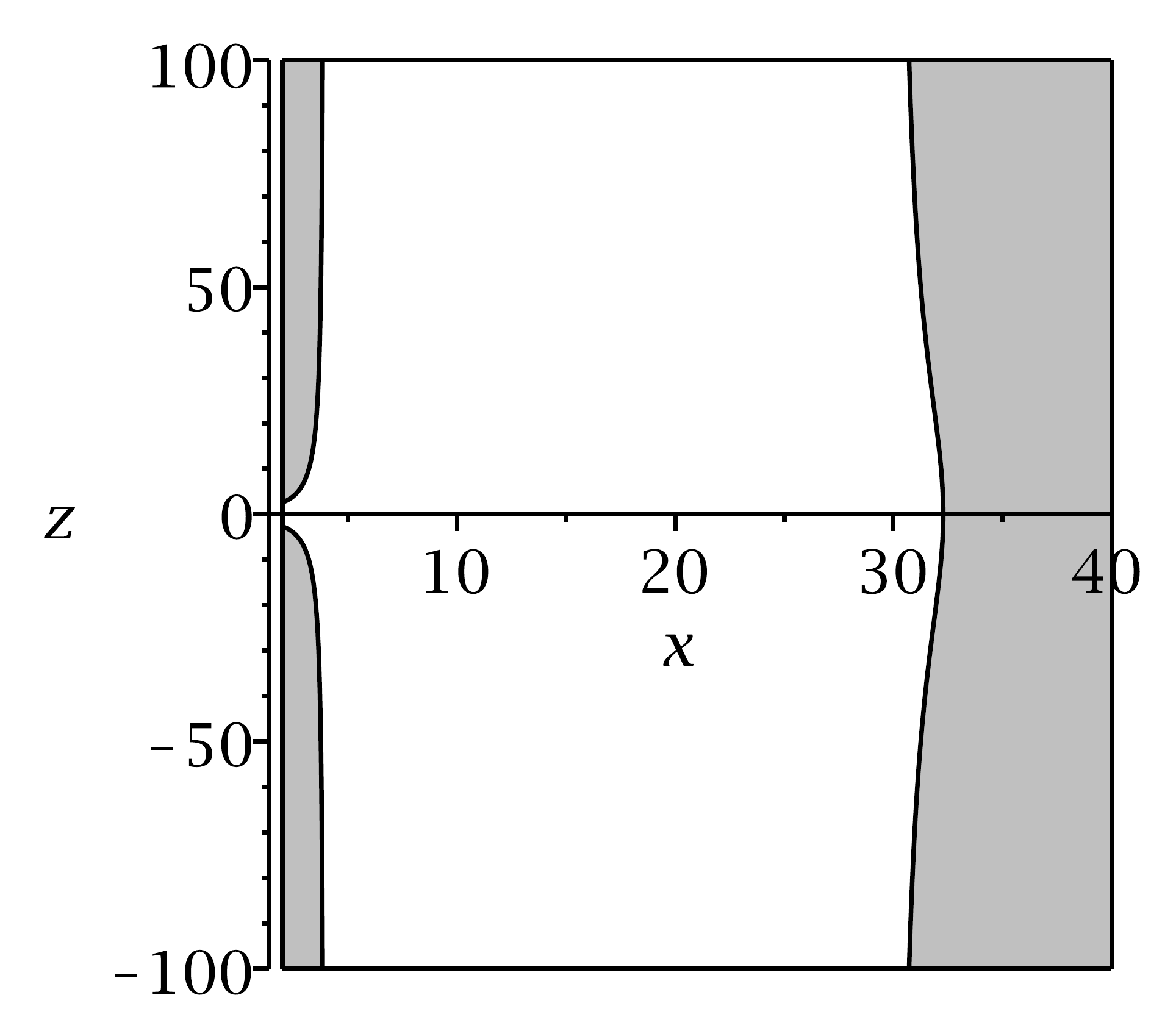}
   \caption{$\Ecal=1.2$.}
   \label{esc-4}
  \end{subfigure}
 \end{center}
 \caption{(Color online) Curves of $\Ucal^2=\Ecal^2$ plotted as the boundary separating the regions accessible (white) and inaccessible (shaded) for a particle kicked from an ISCO into various possible energies. The spacetime parameters are $B=0.005$, $\nu=0.9$ and $r_g=2$, and the charge per unit mass of the particle is $e=10$. The initial ISCO radius for these spacetime parameters is $r_{(\mathrm{ISCO})}=4.9490145$, as can be calculated from \Eqref{circ_perturb}. The corresponding angular momentum is $\Lcal_{-(\mathrm{ISCO})}=2.986451488$.}
 \label{esc}
\end{figure}
\begin{figure}
 \begin{center}
  \begin{subfigure}[b]{0.45\textwidth}
   \centering
   \includegraphics[angle=-90,scale=0.7]{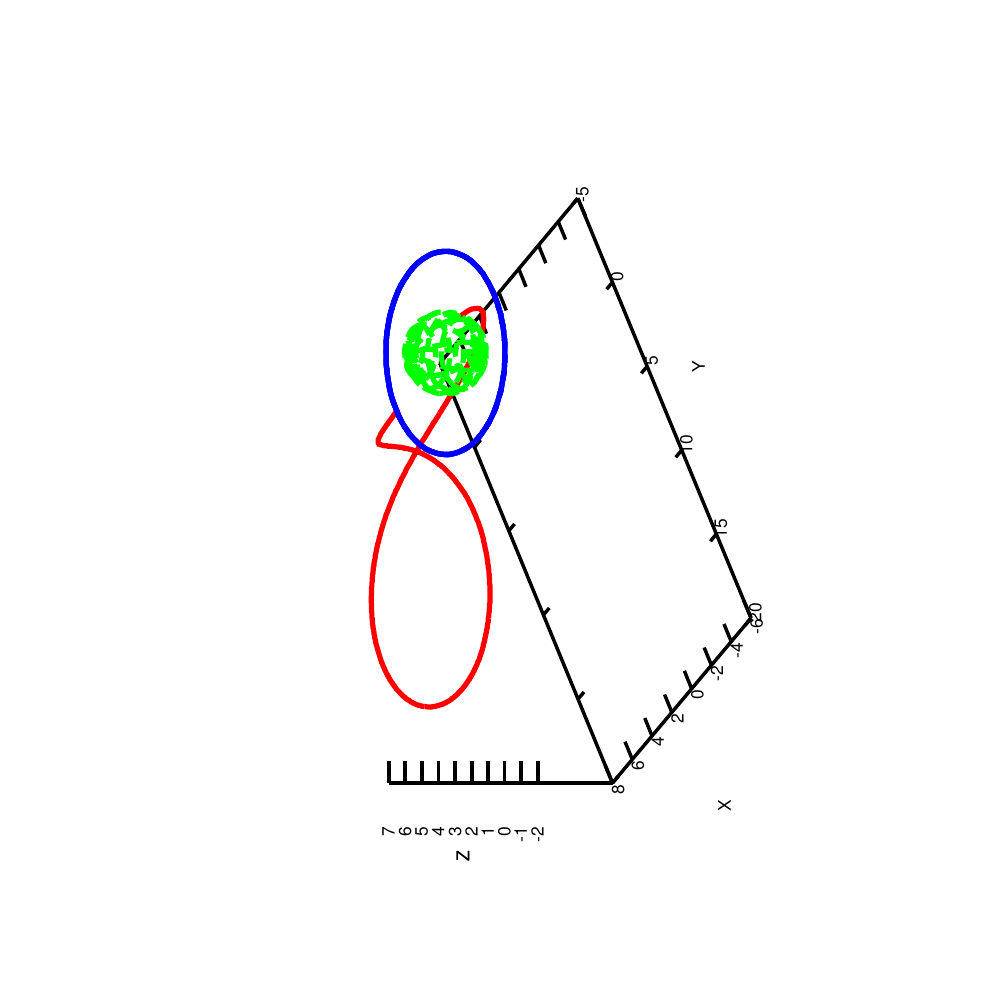}
   \caption{$\Ecal=1.0$.}
   \label{kick-1}
  \end{subfigure}
  \begin{subfigure}[b]{0.45\textwidth}
   \centering
   \includegraphics[angle=-90,scale=0.7]{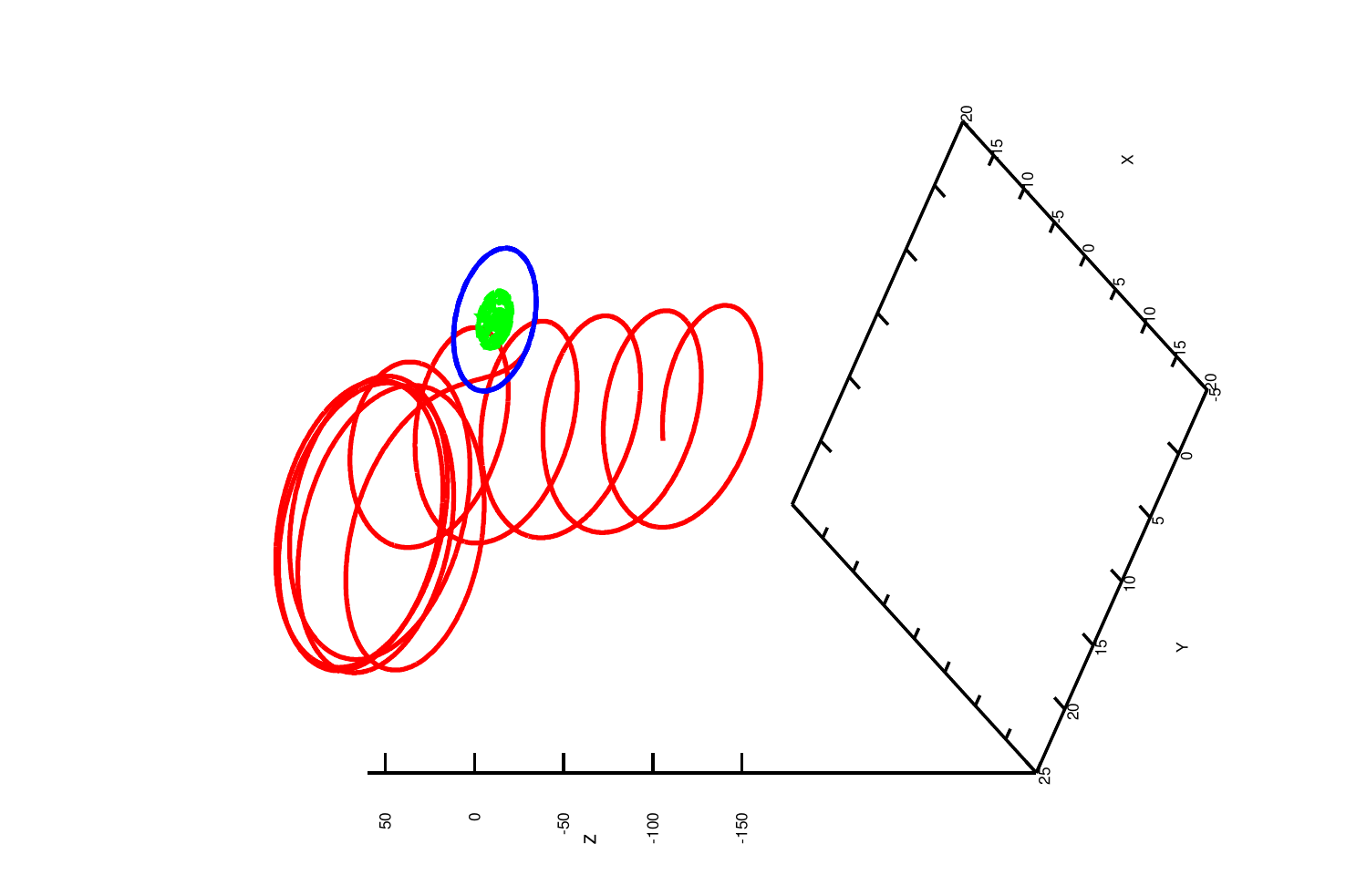}
   \caption{$\Ecal=1.1$.}
   \label{kick-2}
  \end{subfigure}
  \caption{(Color online) Trajectory of a particle kicked from an initial ISCO orbit where $B=0.005$, $\nu=0.9$, $r_g=2$, $e=10$ and $\left|\Lcal_{-(\mathrm{ISCO})}\right|=2.986451488$, for post-collision energies of $\Ecal=1.0$ and $\Ecal=1.1$. The sphere indicates the surface of the singularity $r=r_g=2$, and the blue circles denote the particle's original ISCO orbit ($r_{(\mathrm{ISCO})}=4.9490145$) before the collision.}
  \label{kick}
 \end{center}
\end{figure}

The dark regions in Figs. (\ref{esc-1})--(\ref{esc-4}) indicate
regions inaccessible to the particle. In particular, note that Figs.
(\ref{esc-3}) and (\ref{esc-4}) shows that the particle has the
ability to escape far from the origin if it travels up or down where
$\left| z\right|\gg 1$. Recalling that, at $\phi=0$, we have
$z=r\sin\theta$, hence this is consistent with the observation made
in Eq.~\Eqref{inf_barrier}.
The actual trajectory of a kicked particle may be obtained by
integrating Eqs.~\Eqref{rddot} and \Eqref{thetaddot} numerically.
Continuing the example where $B=0.005$, $\nu=0.9$, $r_g=2$ and
$\Lcal_{-(\mathrm{ISCO})}$, we track the motion of the particle
after being kicked from the initial orbit at $r=r_{(\mathrm{ISCO})}$
into a new energy $\Ecal$, with the angular momentum remaining the
same and $\dot{r}=0$ just after collision. The initial value of
$\dot{\theta}$ just after collision is calculated from
\Eqref{first_integral}. With these initial conditions, we plot the
trajectories for two possible values of $\Ecal=1.0$ and $\Ecal=1.1$,
shown in Fig. (\ref{kick}).
\par
As we can see, the particle after the kick may possibly fall into
the $r=r_g$ singularity, as shown in Fig. (\ref{kick-1}), or escape,
shown in Fig. (\ref{kick-2}) while continuing to execute a
cyclotron-like spiral due to the Lorentz interaction with the
magnetic field.

\section{Escape velocity in the unmagnetized JNW spacetime}\label{jnw}
In this section, we consider the unmagnetized case where $B=0$. The angular momentum and energy representing circular orbits, Eq.~\Eqref{L_circ} and \Eqref{E_circ} for $B=0$ are
\begin{align}
 \Lcal_{o}=r_o\sqrt{\frac{\nu\left(1-\frac{r_g}{r_o}\right)^{1-\nu}r_g}{2r_o-(1+2\nu)r_g}},
\quad\Ecal_o=\sqrt{\left(1-\frac{r_g}{r_o}\right)^\nu\left(\frac{2r_o-r_g(1+\nu)}{2r_o-r_g(1+2\nu)}\right)}.
\end{align}
For $\nu=1/2$,  the orbit exists for $r_o\in(\frac{3}{2}r_g,\infty)$
and the ISCO is specified by $r_o=\frac{3}{2}r_g$, which corresponds
to an inflection point of the effective potential. For the ISCO we
have
\begin{equation}
\mathcal{E}_{ISCO}=\frac{3^{\frac{1}{4}}}{\sqrt{2}}, \quad
|\mathcal{L}_{zISCO}|=\frac{3^\frac{3}{4}}{2\sqrt{2}}r_g.
\end{equation}

As already described in Sec.~\ref{esc_trajectories_mjnw}, we
consider collision of particles initially in an ISCO with another
particle, where we assume $\dot{r}=0$ after the collision and
$\Lcal$ remains unaltered, and the motion is mainly determined by
the energy $\Ecal$ after the collision. Since this implies
$\dot{\theta}\neq0$ after the collision, the particles obtains a
velocity $v_\perp=r\dot{\theta}_o$ ($\dot{\theta}_o$ is the initial
polar angular velocity) in the direction orthogonal to the
equatorial plane. The energy then takes the form
\begin{equation}
\mathcal{E}=\sqrt{\mathcal{E}_o^2+v^2_\bot\left(1-\frac{r_g}{r_o}\right)},
\end{equation}
If $\mathcal{E}<1$ particle cannot escape to infinity, it
will escape to infinity if $\mathcal{E}\geq1$ or in other words it
will have unbounded motion,
\begin{equation}
v_\bot\geq\sqrt{(1-\mathcal{E}^2_o)\left(\frac{r_o}{r_o-r_g}\right)}.
\end{equation}
Specifically, for  ISCO the escape condition for $\nu=\frac{1}{2}$
is, $|v_\bot|\geq
v^{esc}_\bot\geq\sqrt{3\left(1-\frac{\sqrt{3}}{2}\right)}$.
\label{2}

\section{Particles in a weakly magnetized JNW spacetime} \label{wmjnw}

\subsection{Equations of motion in the weak magnetization regime}

If the magnetic field is sufficiently weak such that they do not influence the particle gravitationally, there is a possibility for a particle to escape after collision. This may occur if $\Lambda\rightarrow 1$, or, equivalently, $B\rightarrow 0$. Nevertheless, the magnetic field may still influence the particle via the Lorentz force, as the coupling depends on $eB$. Thus, even for small $B$, the Lorentz force will remain significant for sufficiently large $e$.
\par
Therefore, for timelike particles ($\epsilon=-1$), we shall focus on this weak magnetization regime by defining $B=\beta/e$ and expand the equations of motion \Eqref{rddot} and \Eqref{thetaddot} in powers of $1/e$, the result is
\begin{align}
 \ddot{r}&=\frac{\nu r_g}{2r(r-r_g)}\dot{r}^2+\half\brac{2r-r_g(1+\nu)}\dot{\theta}^2-\frac{\nu r_g\Ecal^2}{2r(r-r_g)}\nonumber\\
          &\quad+\brac{\frac{r-r_g}{r}}^{2\nu}\frac{\sbrac{2r-r_g(1+\nu)}\Lcal^2}{2(r-r_g)^2r^2\sin^2\theta}+\mathcal{O}(1/e^2),\label{rddot_test}\\
 \ddot{\theta}&=-\frac{2r-r_g(1+\nu)}{r(r-r_g)}\dot{r}\dot{\theta}+\brac{\frac{r-r_g}{r}}^{2\nu}\frac{\Lcal^2\cos\theta}{r^2(r-r_g)^2\sin^3\theta}-\frac{\beta^2}{4}\cos\theta\sin\theta+\mathcal{O}(1/e^2), \label{thetaddot_test}
\end{align}
and the constraint equation becomes
\begin{align}
 0=-\frac{\Ecal^2}{F}+\frac{\Lcal^2}{r^2G\sin^2\theta}-\beta\Lcal+\frac{\dot{r}^2}{F}+r^2G\dot{\theta}^2+1+\frac{\beta^2}{4}r^2G\sin^2\theta+\mathcal{O}(1/e^2). \label{first_test}
\end{align}
Substituting Eq.~\Eqref{first_test} into \Eqref{rddot_test}, and expressing the quantities in terms of $f$, we obtain
\begin{align}
 \ddot{r}&=\half\brac{2r-(1+2\nu)r_g}\brac{\dot{\theta}^2+\frac{\Lcal^2}{r^4f^{2(1-\nu)}\sin^2\theta}}+\frac{\nu r_g}{2r^2f^{1-\nu}}\brac{\beta\Lcal - 1}-\frac{\beta^2}{8}(2r-r_g)\sin^2\theta,\label{s6}\\
 \ddot{\theta}&=-\frac{2}{r}\brac{1+(1-\nu)\frac{r_g}{2rf}}\dot{r}\dot{\theta}+\frac{\Lcal^2\cos\theta}{r^4f^{2(1-\nu)}\sin^3\theta}-\frac{\beta^2}{4}\sin\theta\cos\theta.\label{s7}
\end{align}
Accordingly, to first-order in $1/e$, Eq.~\Eqref{tdotphidot} reduces to
\begin{align}
 \Ecal=f^\nu\dot{t},\quad\Lcal=\brac{\dot{\phi}+\frac{\beta}{2}}r^2f^{1-\nu}\sin^2\theta. \label{s8}
\end{align}
Rearranging Eq.~\Eqref{first_test} we obtain an effective potential equation for the test field case
\begin{align}
 \Ecal^2=&\;\dot{r}^2+r^2f\dot{\theta}^2+U_{\mathrm{eff}},\nonumber\\
  U_{\mathrm{eff}}=&\;f^\nu\sbrac{1+r^2f^{(1-\nu)}\sin^2\theta\brac{\frac{\Lcal}{r^2f^{(1-\nu)}\sin^2\theta}-\frac{\beta}{2}}^2}. \label{Ueff_test}
\end{align}

\subsection{Dimensionless form of the equations}
Following dimensionless quantities are introduced to avoid
complications in our analysis regarding motion of the particle after
collision,
\begin{equation}
\sigma=\frac{\tau}{r_g}, \quad \rho=\frac{r}{r_g}, \quad
\ell=\frac{\mathcal{L}_z}{r_g}, \quad b=\half\beta r_g, \quad
\mathcal{T}=\frac{t}{r_g}.
\end{equation}
The $\rho$ and $\theta$ components of the dynamical equations
(\ref{s6}) and (\ref{s7})  are expressed as follows:
\begin{equation}
\begin{split}
\frac{d^2\rho}{d\sigma^2}&=\frac{1}{2}(2\rho-(1+2\nu))\left(\frac{d\theta}{d\sigma}\right)^2+\frac{\ell^2(2\rho-(1+2\nu))}{2\rho^4\sin^2\theta
\left(1-\frac{1}{\rho}\right)^{2(1-\nu)}}+ \\
&\frac{\nu}{2\rho^2\left(1-\frac{1}{\rho}\right)^{(1-\nu)}}(2\ell
 b-1) -\frac{b^2}{2}\sin^2\theta(2\rho-1),\label{GG1}
\end{split}
\end{equation}
\begin{equation}
\begin{split}
\frac{d^2\theta}{d\sigma^2}&=-\frac{2}{\rho}\frac{d\theta}{d\sigma}\frac{d\rho}{d\sigma}-\frac{(1-\nu)}{\rho^2\left(1-\frac{1}{\rho}\right)}\frac{d\theta}{d\sigma}\frac{d\rho}{d\sigma}
+\frac{\ell^2\cos\theta}{\rho^4\sin^3\theta\left(
1-\frac{1}{\rho}\right)^{2(1-\nu)}} \\  &-b^2\sin\theta
\cos\theta,\label{GG2}
\end{split}
\end{equation}
where the energy $\mathcal{E}$ is
\begin{equation}\label{Fcircle}
\mathcal{E}^2=\left(\frac{d\rho}{d\sigma}\right)^2+\rho(\rho-1)\left(\frac{d\theta}{d\sigma}\right)^2+U_{\textrm{eff}},
\end{equation}
and
\begin{equation}\label{s14}
U_{\textrm{eff}}=\left(1-\frac{1}{\rho}\right)^\nu+\rho^2\left(1-\frac{1}{\rho}\right)\sin^2\theta\left(\frac{\ell}{\rho^2\sin^2\theta
\left(1-\frac{1}{\rho}\right)^{1-\nu}}-b\right)^2.
\end{equation}
When we consider the equatorial plane, then the particle moving
around the naked singularity in the ISCO has the following energy,
\begin{equation}
\mathcal{E}^2_o=\left(1-\frac{1}{\rho}_o\right)^\nu+\rho^2_o\left(1-\frac{1}{\rho}_o\right)\left(\frac{\ell}{\rho^2_o\left(1-\frac{1}{\rho}_o\right)^{1-\nu}}-b\right)^2.
\end{equation}
After the collision the motion of particle is merely determined by
its new  energy
\begin{equation}\label{Ecircle}
\mathcal{E}=\sqrt{{\mathcal{E}^2_o}+\frac{(\rho_o-1)}{\rho_o}v^2_\bot}.
\end{equation}
The particle attains a velocity after the collision i.e. orthogonal
to the equatorial plane given as
\begin{align}\label{Gcircle}
 v_{\bot}=\sqrt{(\mathcal{E}^2-\mathcal{E}^2_o)\frac{\rho_o}{\rho_o-1}}.
\end{align}
We examine the motion of the particle after collision, initially
orbiting in the ISCO.  The dimensionless forms of the angular
momentum and magnetic field represented as $\ell$ and $b$,
respectively are given below \cite{Frolov:2010mi}:
$\frac{dU}{d\rho}=U'=0$ and $\frac{d^2U}{d\rho^2}=U''=0$ are used to
find $\ell$ and $b$,
\begin{equation}
\begin{split}\label{u1}
U_{\textrm{eff}}'&=\frac{1}{\rho^2(\rho-1)^2}[-\ell^2(1-\frac{1}{\rho})^{2\nu}(2\rho-1-2\nu)-\\
&2\ell b(\rho-1)(1-\frac{1}{\rho})^{\nu}\rho\nu
+\rho(\rho-1)(b^2\rho(1-3\rho+2\rho^2) + \\
&\nu(1-\frac{1}{\rho})^\nu)],
\end{split}
\end{equation}

\begin{equation}\label{u2}
\begin{split}
U_{\textrm{eff}}''&=\frac{1}{\rho^3(\rho-1)^3}[2lb(\rho-1)(1-\frac{1}{\rho})^\nu\rho\nu(2\rho-1-\nu)\\
&+\rho(\rho-1)(2b^2\rho^2(\rho-1)^2
+(1-\frac{1}{\rho})^\nu\nu(1-2\rho+\nu))\\&+2\ell^2(1-\frac{1}{\rho})^{2\nu}(1+3\rho^2+3\nu+2\nu^2-3\rho(1+2\nu))].
\end{split}
\end{equation}
Equating Eq. (\ref{u1}) and Eq. (\ref{u2}) to zero, and solving them
simultaneously gives
\begin{align}\label{s12}
\ell=&\pm\sqrt{\frac{-b^2(\rho_o-1)^2\rho_o^2(1+6\rho_o^2+\nu-2\rho_o(3+\nu))}{(1-\frac{1}{\rho_o})^{2\nu}(1+2\rho_o^2+3\nu+2\nu^2-2\rho_o(1+3\nu))}},
\\
\label{s13}
b=&\sqrt{{\frac{-(1-\frac{1}{\rho_o})^{\nu-1}\nu(1+2\rho_o^2+3\nu+2\nu^2-2\rho_o(1+3\nu))}{2\rho_o^2(-1+8\rho_o^3-3\nu-2\nu^2-2\rho_o^2(6+7\nu)+2\rho_o(3+7\nu+2\nu^2)\pm\nu\sqrt{\Sigma})}}},
\end{align}

where
\begin{equation}
\begin{split}
\Sigma&=-[12\rho_o^4+(1+\nu)^2(1+2\nu)-8\rho_o^3(3+5\nu)+4\rho_o^2(5+15\nu+6\nu^2)\\
&-4\rho_o(2+7\nu+6\nu^2+\nu^3)].
\end{split}
\end{equation}
\label{5}

For $\nu=0.8$, when $\ell>0$  $\rho_o \in(1,2.44]$ and when $\ell<0$
$\rho_o \in [1.8,2.44]$. Also, for $\nu=0.9$, when $\ell>0$
 $\rho_o \in(1,2.72]$ and for $\ell<0$  $\rho_o \in[2,2.72]$

\begin{figure}[!ht]
\centering
\begin{minipage}[!ht]{.45\textwidth}
\includegraphics[width=\textwidth]{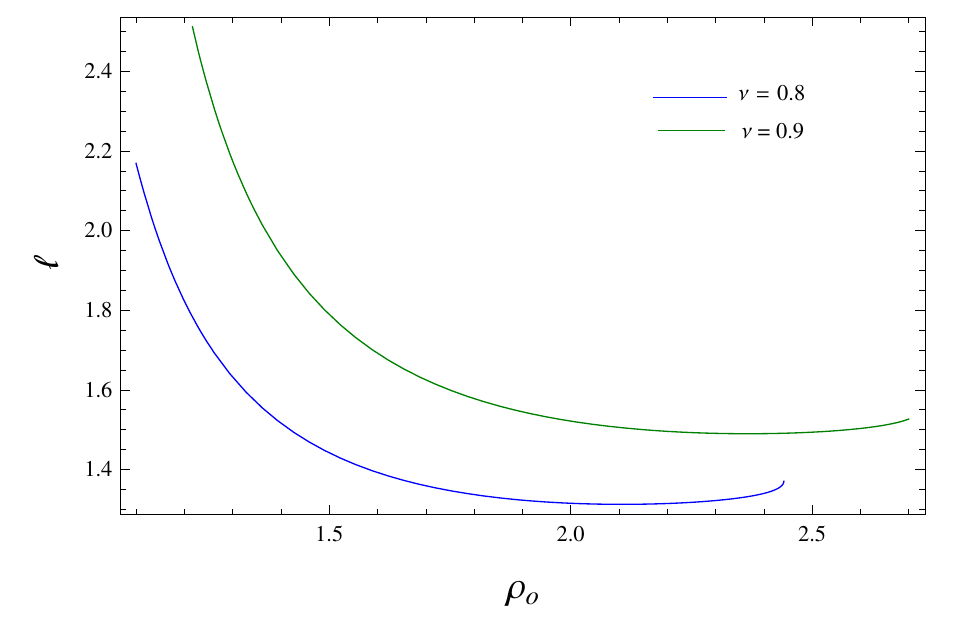}\caption{Angular momentum as a function of radius when $\ell>0$.}\label{kk28}
\end{minipage} \hfill
\begin{minipage}[!ht]{.45\textwidth}
\includegraphics[width=\textwidth]{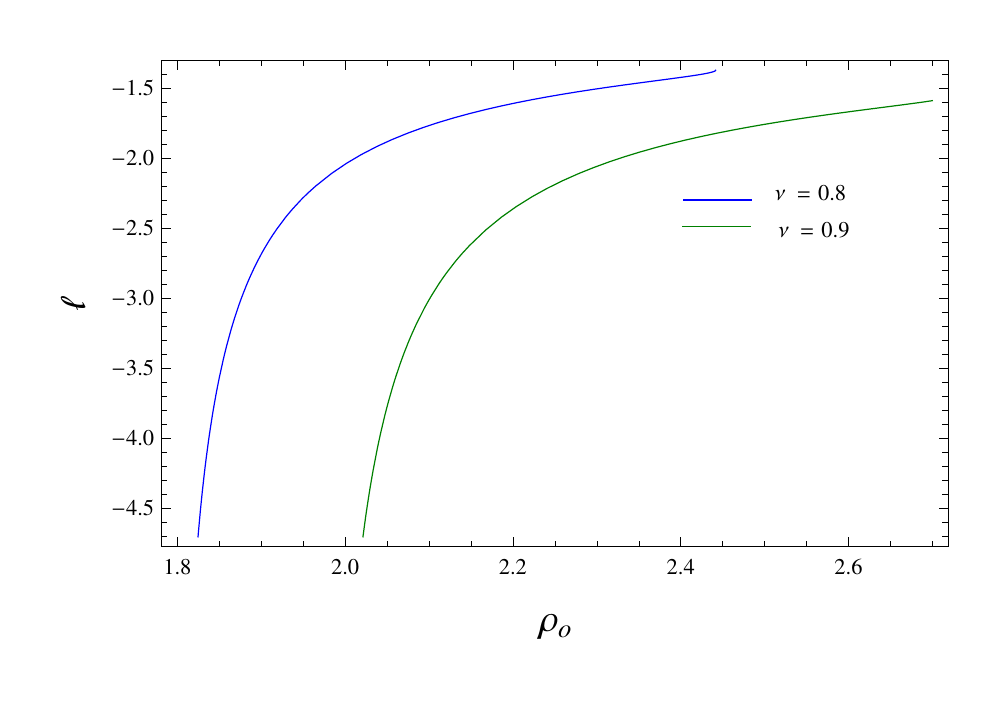}
\caption{Angular momentum as a function of radius when $\ell<0$.
}\label{kk29}
\end{minipage}
\end{figure}

\begin{figure}[!ht]
\centering
\begin{minipage}[!ht]{.45\textwidth}
\includegraphics[width=\textwidth]{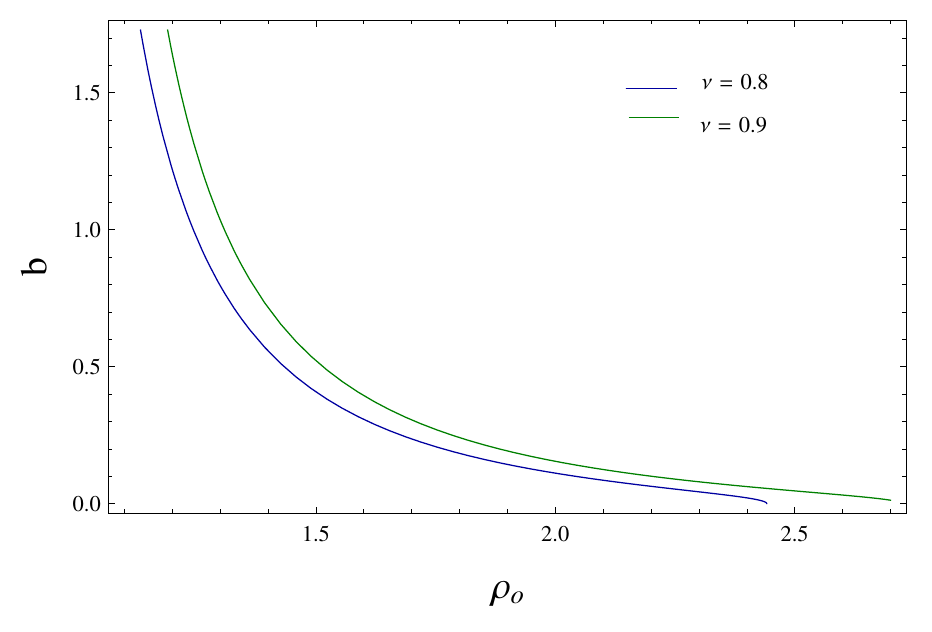}\caption{Magnetic field as a function of radius when $\ell>0$.}\label{kk30}
\end{minipage} \hfill
\begin{minipage}[!ht]{.45\textwidth}
\includegraphics[width=\textwidth]{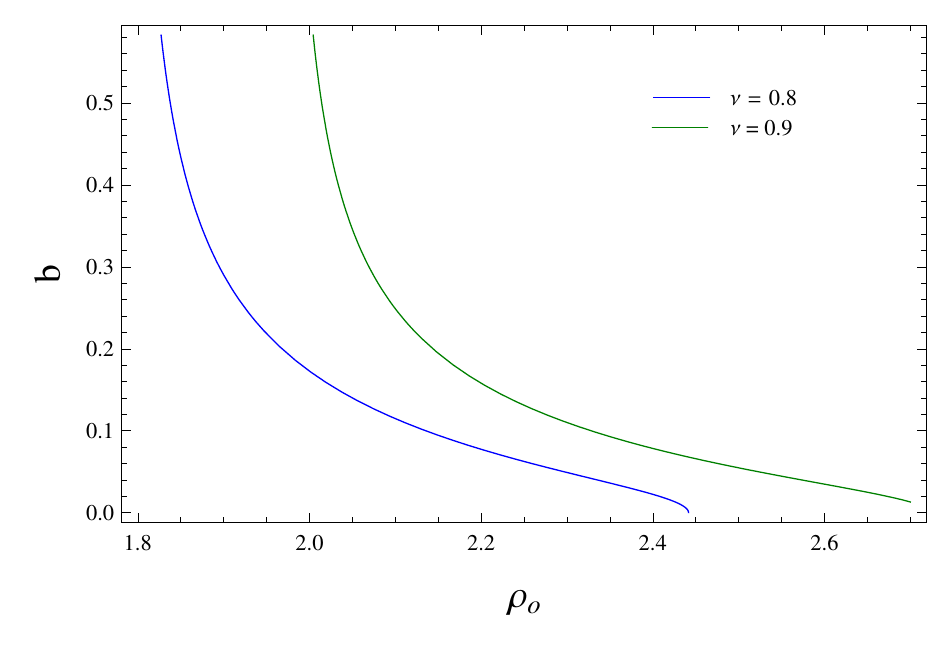}
\caption{Magnetic field as a function of radius when $\ell<0$.
}\label{kk31}
\end{minipage}
\end{figure}

In Fig. (\ref{kk28}) it is shown that as the particle moves away
from the naked singularity the positive angular momentum decreases.
In Fig. (\ref{kk29}) the negative angular momentum increases as a
function of $\rho_o$. In Fig. (\ref{kk30}) and Fig. (\ref{kk31}) it
is evident that the magnetic field strength decreases as radius of
the orbit increases.

Considering the equatorial plane and utilizing the above mentioned
dimensionless quantities Eqs. (\ref{s8}) and (\ref{Ueff_test}) take
the form,

\begin{equation}
 (\frac{d\rho_o}{d\sigma})^2=\mathcal{E}^2-U_{\textrm{eff}},\quad
 \frac{d\mathcal{T}}{d\sigma}=\frac{\mathcal{E}}{(1-\frac{1}{\rho_o})^\nu},
 \end{equation}
 \begin{equation}
 \rho_o\frac{d\phi}{d\sigma}=\alpha, \quad
 \alpha=\frac{\ell}{\rho_o(1-\frac{1}{\rho_o})^{1-\nu}}-b\rho_o.
 \end{equation}

 The effective potential  given below is same as in Eq. (\ref{s14})
 in the equatorial plane but, here we introduce $\alpha$ which is later used to
 define the Lorentz gamma factor

\begin{equation}
U_{\textrm{eff}}=\left(1-\frac{1}{\rho_o}\right)^\nu+\left(1-\frac{1}{\rho_o}\right)\alpha^2.
\end{equation}

 We consider a positive charged particle, so that $b$ is positive as
 well. Let us suppose that the particle is moving in the circular orbit
 of radius $r$. Its momentum is given by

\begin{equation}
 k^\mu=m\gamma (e^\mu_{(t)}+\upsilon e^\mu_{(\phi)}),
\end{equation}
\begin{align}\label{ww1}
e^\mu_{(t)}=&f^{-\nu/2}\xi^\mu_{(t)},\\ \label{ww2}
 e^\mu_{(\phi)}=&r^{-1}
f^{-(1-\nu)/2}\xi^\mu_{(\phi)}.
\end{align}

The velocity of the particle  with respect to a rest frame is
represented by  $v$. Here, $\gamma$ is the Lorentz gamma factor.
Using the normalization condition  $k^2=-m^2$ we obtain
$\gamma=(1-v^2)^{-1/2}$. In Eqs. (\ref{ww1},\ref{ww2})
 $\xi^\mu_{(t)}=(1,0,0,0)$ and $\xi^\mu_{(\phi)}=(0,0,0,1)$ are the
Killing vectors whereas $e^\mu_{(t)}$ and $e^\mu_{(\phi)}$  refers
to the tetrad basis. For a positively charged particle and $v>0$,
the Lorentz force  is $repulsive$ (i.e. directed outwards the naked
singularity), while for $v<0$,  it is $attractive$.

Using relation $d\phi/d\tau=v\gamma/r$ one gets
\begin{equation}
\upsilon\gamma=\alpha.
\end{equation}

This relation allows one to write
\begin{equation}
\gamma^2=1+\alpha^2, \quad v=\frac{\alpha}{\sqrt{1+\alpha^2}}.
\end{equation}

Now we discuss the  behavior of $v$, effective potential and the
Lorentz gamma factor for $\nu=0.8$ and $\nu=0.9$.

\begin{figure}[!ht]
\centering
\begin{minipage}[!ht]{.45\textwidth}
\includegraphics[width=\textwidth]{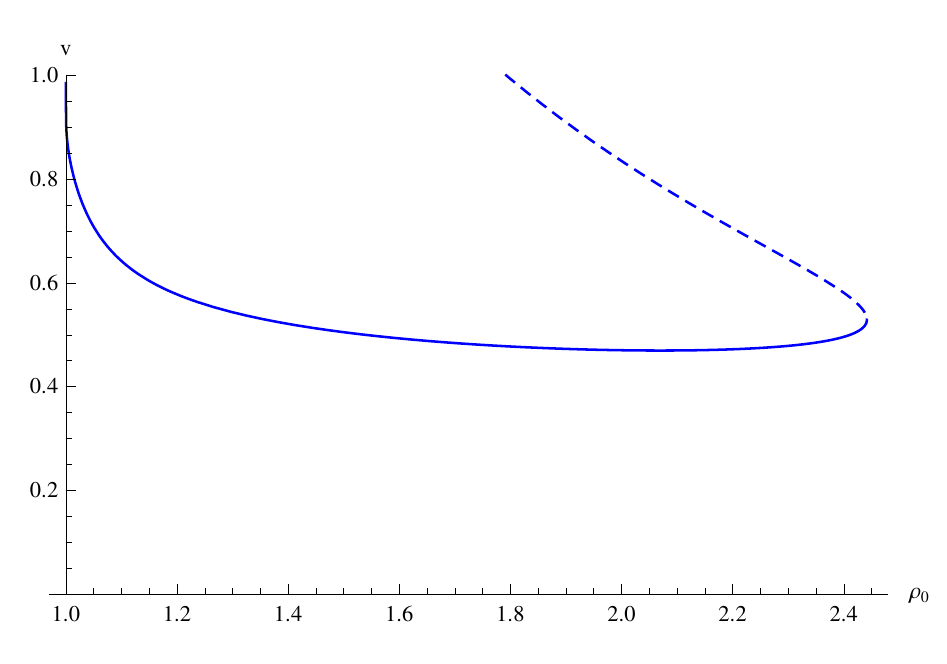}\caption{Velocity with respect to a rest frame for $\nu=0.8$. Dotted curve is for $\ell<0$ and
solid curve  for $\ell>0$.}\label{kk25}
\end{minipage} \hfill
\begin{minipage}[!ht]{.45\textwidth}
\includegraphics[width=\textwidth]{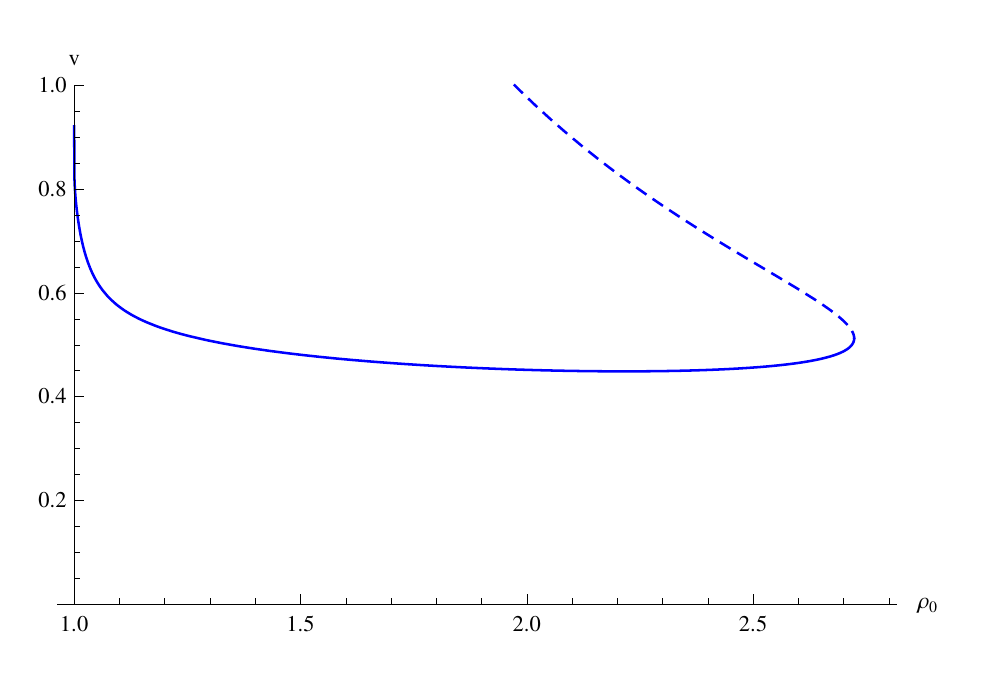}
\caption{Velocity with respect to a rest frame for $\nu=0.9$. Dotted
curve is for  $\ell<0$ and solid curve for $\ell>0$. }\label{kk26}
\end{minipage}
\end{figure}

\begin{figure}[!ht]
\centering
\begin{minipage}[!ht]{.45\textwidth}
\includegraphics[width=\textwidth]{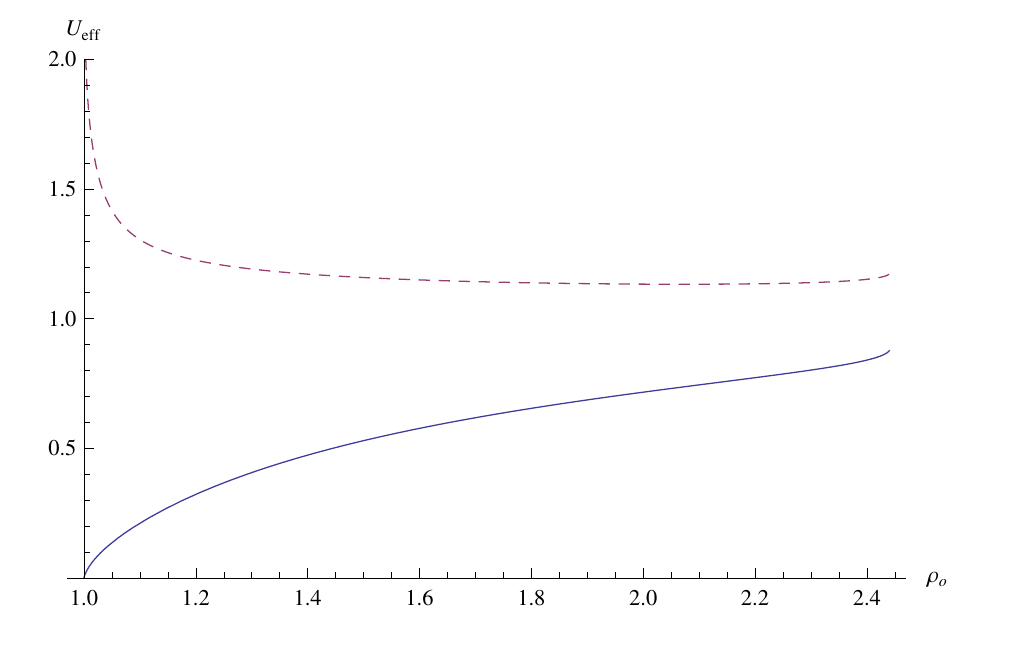}\caption{For $\ell>0$ and $\nu=0.8$, the Lorentz gamma factor (dotted graph) and $U_\textbf{eff}$
are plotted as a function of $\rho_o$.}\label{kk32}
\end{minipage} \hfill
\begin{minipage}[!ht]{.45\textwidth}
\includegraphics[width=\textwidth]{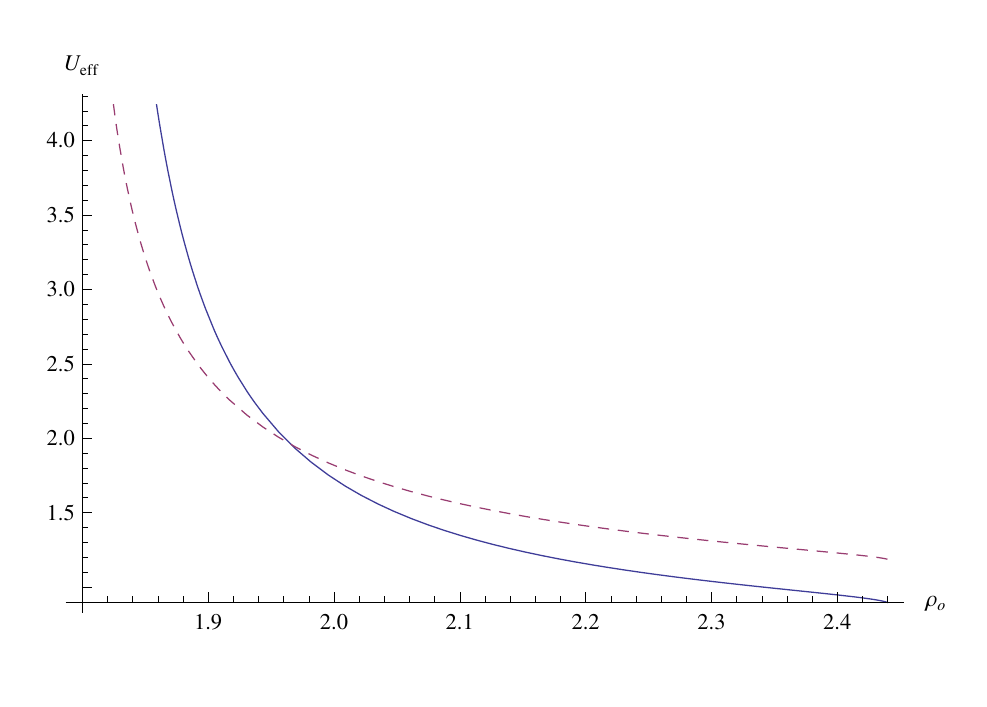}
\caption{For $\ell<0$ and $\nu=0.8$, the Lorentz gamma factor
(dotted graph) and $U_\textbf{eff}$ are plotted as a function of
$\rho_o$. }\label{kk33}
\end{minipage}
\end{figure}

\begin{figure}[!ht]
\centering
\begin{minipage}[!ht]{.45\textwidth}
\includegraphics[width=\textwidth]{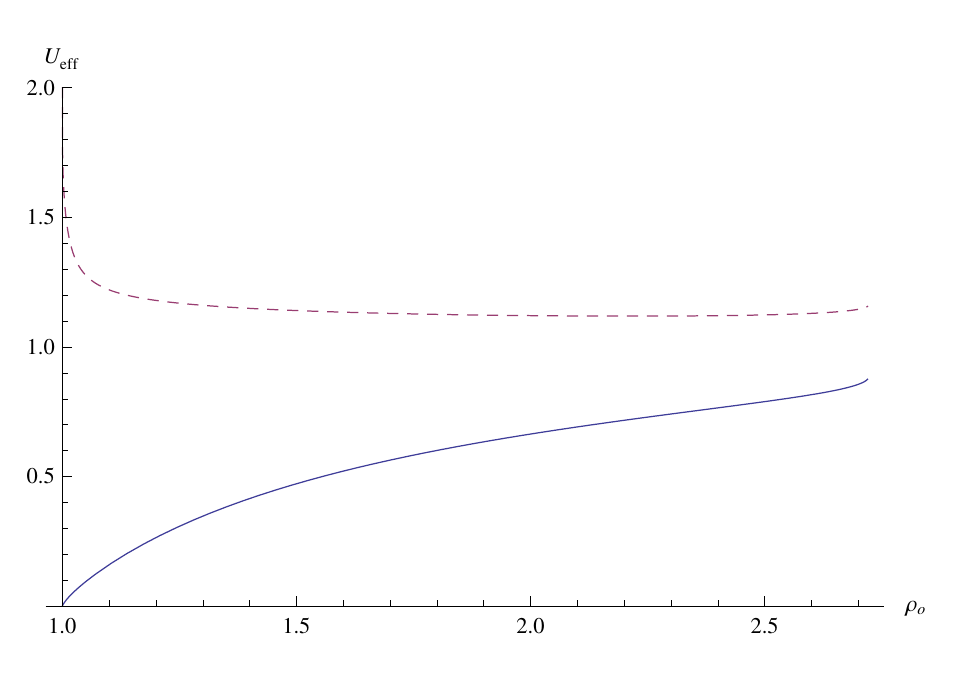}\caption{For $\ell>0$ and $\nu=0.9$, the Lorentz gamma factor (dotted graph) and $U_\textbf{eff}$ as a fuction of $\rho_o$.}\label{kk34}\end{minipage} \hfill
\begin{minipage}[!ht]{.45\textwidth}
\includegraphics[width=\textwidth]{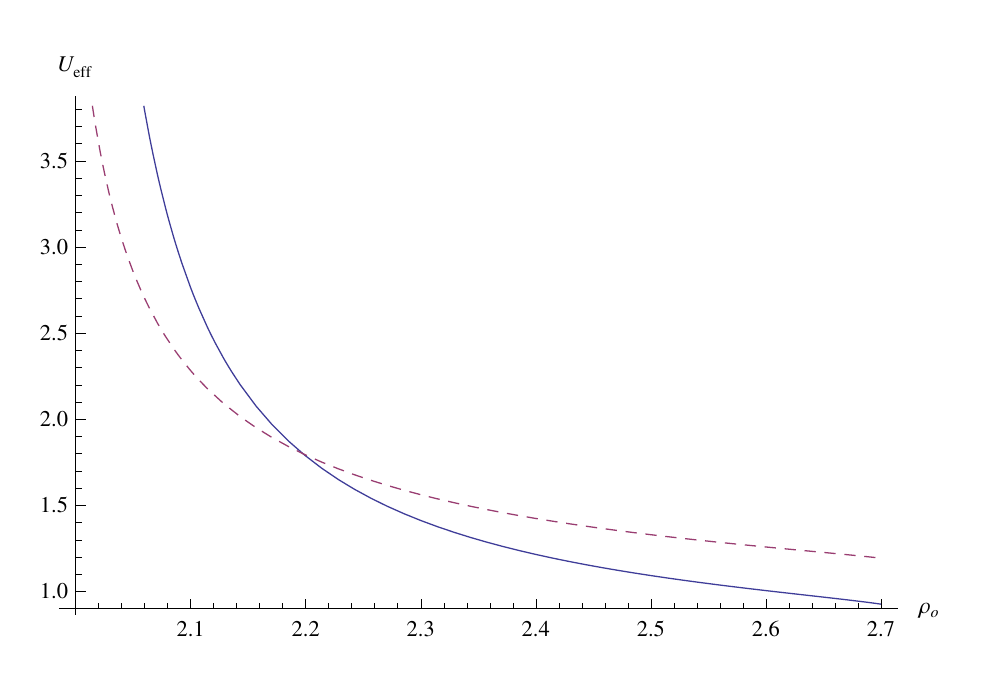}
\caption{For $\ell<0$ and $\nu=0.9$, the Lorentz gamma factor
(dotted graph) and $U_\textbf{eff}$ as a function of $\rho_o$.
}\label{kk35}
\end{minipage}
\end{figure}

The velocity of the particle with respect to a rest frame as a
function of $\rho_o$ is shown in Figs. (\ref{kk25}) and
(\ref{kk26}), $\nu$ = 0.8 and 0.9. It is evident that $v$ decreases
as the particle moves away from the naked singularity, both for
$\ell>0$ and $\ell<0$.
 In Fig. (\ref{kk32}) when $\ell>0$ and $\nu=0.8$,
$U_{\textrm{eff}}$ monotonically decreases from its value $0.88$ at
$\rho_o=2.44$ till it reaches $\rho_o=1$. The Lorentz gamma factor
$\gamma$ at $\rho_o=2.44$ is 1.2. In Fig. (\ref{kk33}) for $\ell<0$
and $\nu=0.8$ both the functions rise to infinity at $\rho_o=1.8$.
In Fig. (\ref{kk34}) considering $\ell>0$ and $\nu=0.9$,
$U_\textbf{eff}$ and $\gamma$ exhibit the same behavior as described
above for $\nu=0.8$, but for different values of $\rho_o$. Both the
functions infinitely grow at $\rho_o=2$ when $\ell<0$ as shown in
Fig. (\ref{kk35}).

\subsubsection{ Energy of the particle at ISCOs}
\begin{figure}[!ht]
\centering
\begin{minipage}[!ht]{.45\textwidth}
\includegraphics[width=\textwidth]{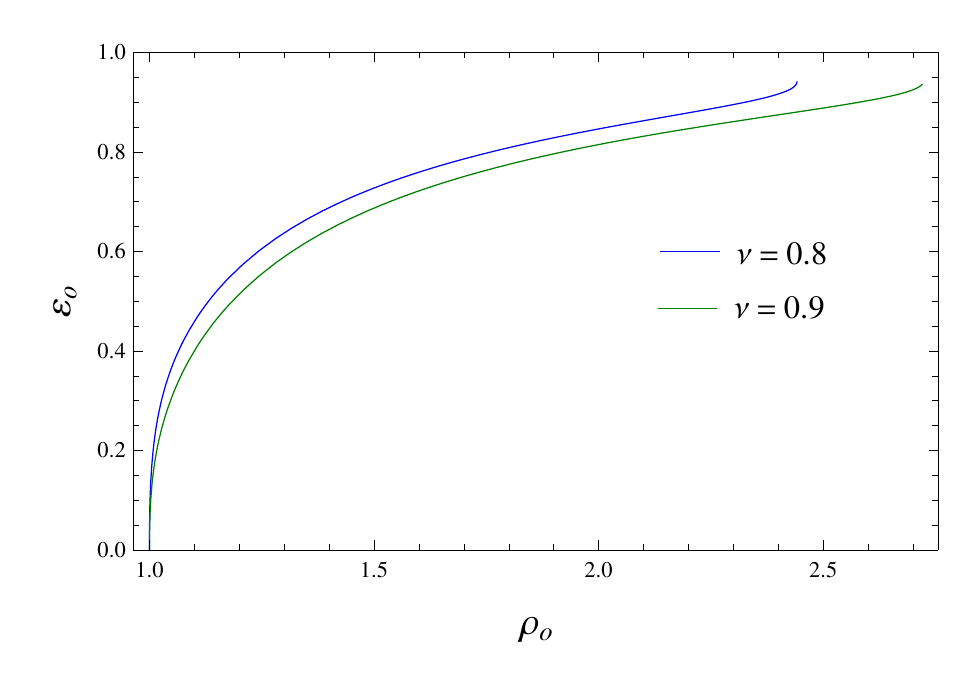}\caption{ Energy of the particle for $\ell>0$.}\label{kk36}
\end{minipage} \hfill
\begin{minipage}[!ht]{.45\textwidth}
\includegraphics[width=\textwidth]{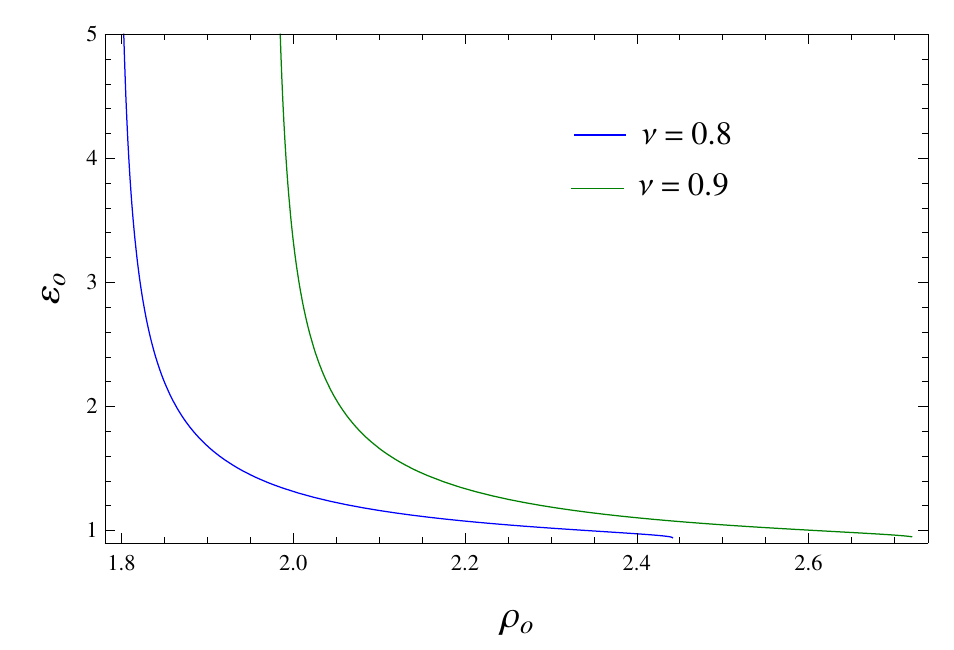}
\caption{Energy of the particle for $\ell<0$. }\label{kk37}
\end{minipage}
\end{figure}

The energy of the particle present in the ISCO for different values
of $\nu$ is demonstrated in Figs. (\ref{kk36},\ref{kk37}). If the
Lorentz force acting on the particle is repulsive i.e., $\ell>0$,
then the energy possessed by the particle is minimum  as compared to
the particle experiencing an attractive Lorentz force i.e.,
$\ell<0$. It is also observed that with the increasing value of
$\nu$, the energy of the particle decreases when $\ell>0$ and
increases when $\ell<0$.

\subsubsection{Escape trajectories and escape velocity }
\begin{figure}[!ht]
\centering
\begin{minipage}[!ht]{.45\textwidth}
\includegraphics[width=\textwidth]{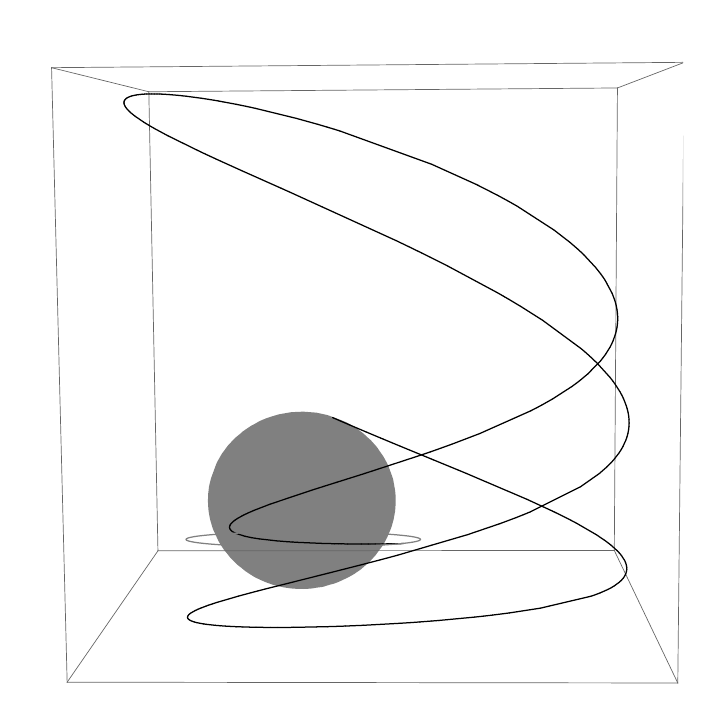}\caption{Escape trajectory for $\nu=0.9$, $\ell=1.49355$, $b=0.0464893$, $\rho_o=2.5$ and $\mathcal{E}=1.16884$.}\label{kk42}
\end{minipage} \hfill
\begin{minipage}[!ht]{.45\textwidth}
\includegraphics[width=\textwidth]{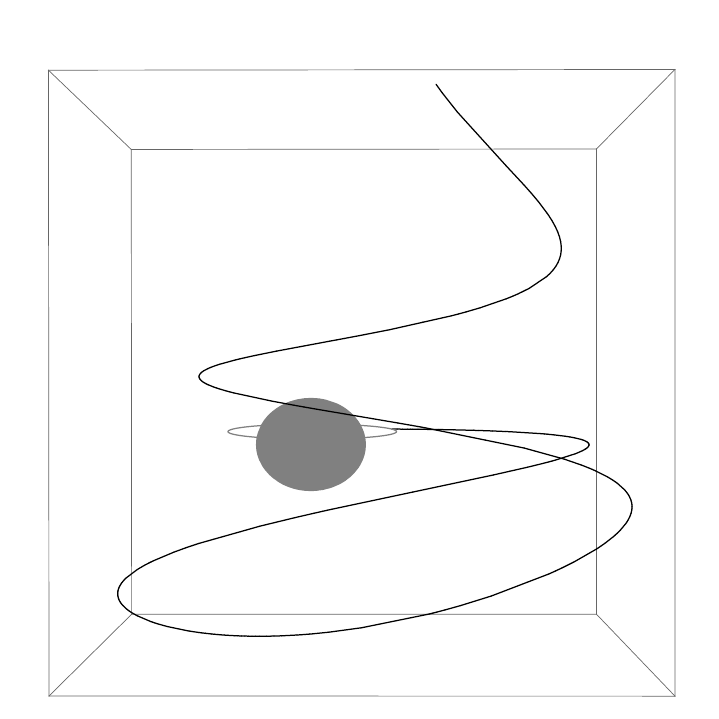}
\caption{Escape trajectory for $\nu=0.9$, $\ell=1.50555$,
$b=0.124502$, $\rho_o=2.1$ and $\mathcal{E}=1.16971$. }\label{kk43}
\end{minipage}
\end{figure}

\begin{figure}[!ht]
\centering
\begin{minipage}[!ht]{.45\textwidth}
\includegraphics[width=\textwidth]{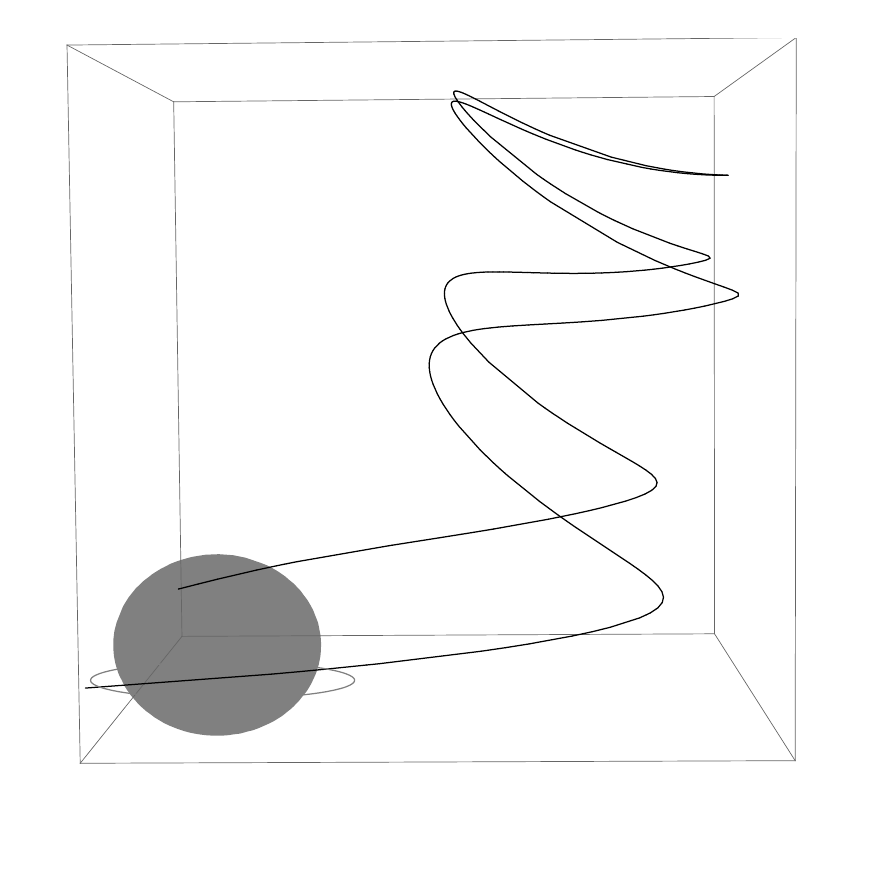}\caption{ Escape trajectory for $\nu=0.9$, $\ell=-1.75741$,
$b=0.0547023$, $\rho_o=2.5$ and $\mathcal{E}=1.082$.}\label{kk44}
\end{minipage} \hfill
\begin{minipage}[!ht]{.45\textwidth}
\includegraphics[width=\textwidth]{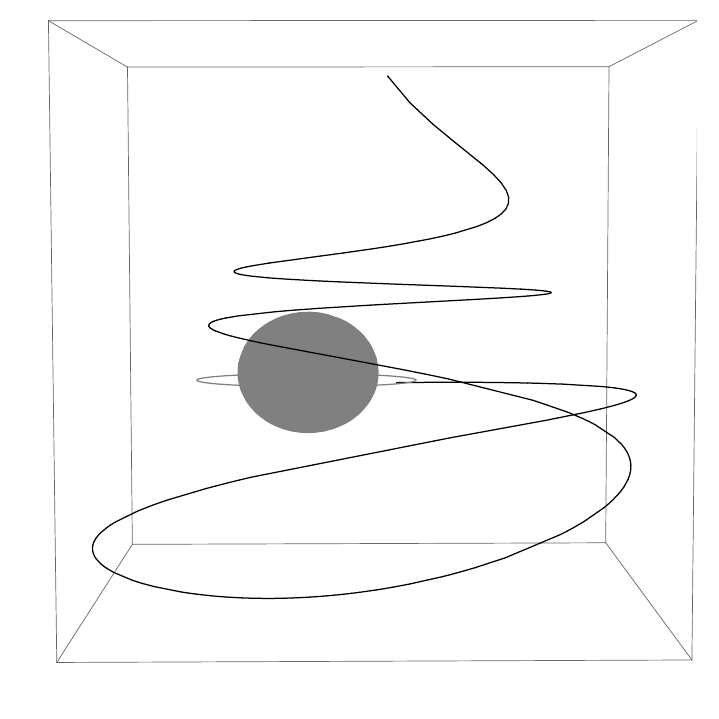}
\caption{Escape trajectory for $\nu=0.9$, $\ell=-2.05737$,
$b=0.109975$, $\rho_o=2.3$ and $\mathcal{E}=1.19955$.}\label{kk45}
\end{minipage}
\end{figure}

\begin{figure}[!ht]
\centering
\begin{minipage}[!ht]{.45\textwidth}
\includegraphics[width=\textwidth]{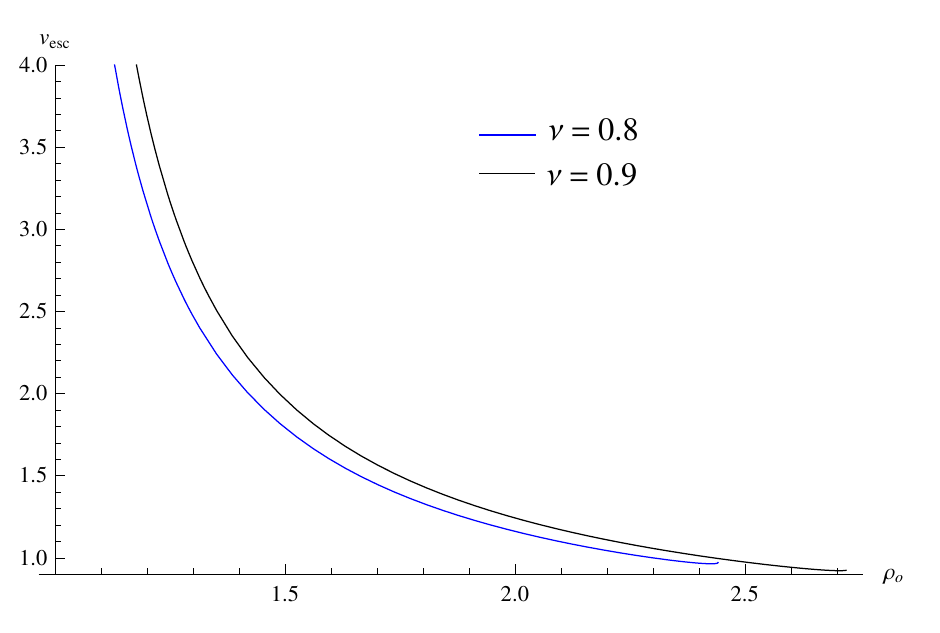}\caption{Escape velocity for $\ell>0$.}\label{kk40}
\end{minipage} \hfill
\begin{minipage}[!ht]{.45\textwidth}
\includegraphics[width=\textwidth]{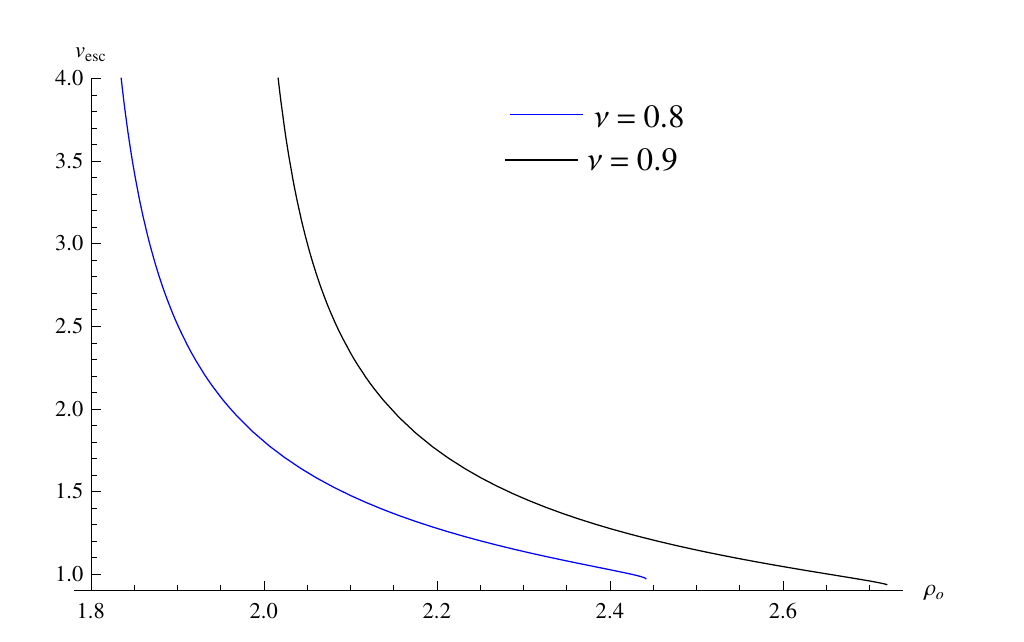}
\caption{Escape velocity for $\ell<0$.}\label{kk41}
\end{minipage}
\end{figure}

 In order to visualize the escape trajectories of the ejected particle
initially present in the ISCO, Eqs. (\ref{GG1},\ref{GG2}) are
numerically solved in Mathematica using NDSolve command by keeping
the radial velocity zero after collision. Figs.
(\ref{kk43},\ref{kk45}) display the path followed by the particle
after being kicked from the ISCO, that ultimately escapes to
infinity. Figs. (\ref{kk42},\ref{kk44}) show capture by the naked
singularity after the collision. Utilizing the Eqs.
(\ref{Fcircle}-\ref{Gcircle}) the escape velocity is estimated from
the numerical results, obtained from the numerical integration of
the dynamical equations \cite{Zahrani:2013up}.  The escape velocity
for different values of $\nu$ is illustrated in Figs.
(\ref{kk40},\ref{kk41}). With the increasing value of $\nu$ the
escape velocity of the particle increases.

\newpage
\section{Particle collision at ISCO} \label{collision}
Let us consider a collision between two particles with the same mass
$m$ and opposite charges, moving along the same circular orbit in
the opposite directions. The four-momentum of the particles is given
as under,
\begin{equation}
k_1^\mu=m\gamma(e^\mu_{(t)}+\upsilon e^\mu_{(\phi)}), \quad
k_2^\mu=m\gamma(e^\mu_{(t)}-\upsilon e^\mu_{(\phi)}),
\end{equation}
\begin{equation}
K^\mu=k_1^\mu+k_2^\mu=2m\gamma e^\mu_{(t)}.
\end{equation}
$K^\mu$ is the four-momentum of the system after the collision. Let
us denote the center-of-mass energy after the collision by
$\mathcal{M}$,
\begin{equation}
\mathcal{M}^2=-K^\mu K_{\mu}=-g_{\mu\delta}K^\mu
K^\delta=(2m\gamma)^2, \quad \mathcal{M}=2m\gamma.
\end{equation}
It is obvious that the center-of-mass energy  $\mathcal{M}$ depends
on the Lorentz gamma factor, as the value of $\gamma$ increases the
energy  generated as a result of the collision is maximum. In Fig.
(\ref{kk32}) when $\ell>0$ and $\nu=0.8$, $\gamma$ remains close to
1. In this case, the energy produced is not significantly large. In
Fig. (\ref{kk33}) when $\ell<0$, $\gamma$ rises to infinity at
$\rho_o$=1.8. Hence, a large amount of energy can be obtained in the
presence of a weak magnetic field if the particle experiences an
attractive Lorentz force. We can also examine the energy generated
for $\nu=0.9$. It exhibits the same behavior as for $\nu=0.8$, but
for different values of $\rho_o$.

.

\section{Conclusion}
In this paper, we established an analysis to investigate the escape
velocity of the particle initially orbiting in the ISCO of a
magnetized JNW naked singularity. The escape trajectories followed
by the particle after the collision are analyzed utilizing the
dynamical equations. The particle may either escape to infinity,
captured by the naked singularity or will continue a bounded motion
after the collision. Also, the center of mass energy is obtained as
a result of the collision between two particles having the same mass
and opposite charges, moving along the same circular orbit in the
opposite directions. It is observed that the maximum amount of
energy can be achieved depending upon the Lorentz gamma factor.

\bibliographystyle{mybib}

\bibliography{jnwrefs}

\end{document}